\PassOptionsToPackage{dvipsnames,svgnames,x11names}{xcolor} 

\documentclass[acmsmall,screen,authorversion]{acmart}

\usepackage{amsmath}
\usepackage{subcaption}
\usepackage{supertabular}
\usepackage{stmaryrd}
\usepackage{xcolor}
\usepackage{multirow}
\usepackage{calc}
\usepackage[T1]{fontenc}
\usepackage{graphicx}
\usepackage{xspace}
\usepackage{dashrule}  
\usepackage{stackrel}
\usepackage{enumerate}
\usepackage{framed}
\usepackage{bussproofs}
\usepackage{mdframed}  
\usepackage{comment}
\usepackage{ifthen}
\usepackage{pdftexcmds}
\usepackage{mathpartir}
\usepackage{fancyvrb}
\usepackage{booktabs}
\usepackage{tikz}
\usepackage{tikz-cd}
\usepackage[zi4]{inconsolata-slanted} 
\usepackage{listings}
\usepackage{url}
\usepackage{float}
\usepackage{subfiles}

\usetikzlibrary{calc}
\tikzset{curve/.style={settings={#1},to path={(\tikztostart)
    .. controls ($(\tikztostart)!\pv{pos}!(\tikztotarget)!\pv{height}!270:(\tikztotarget)$)
    and ($(\tikztostart)!1-\pv{pos}!(\tikztotarget)!\pv{height}!270:(\tikztotarget)$)
    .. (\tikztotarget)\tikztonodes}},
    settings/.code={\tikzset{quiver/.cd,#1}
        \def\pv##1{\pgfkeysvalueof{/tikz/quiver/##1}}},
    quiver/.cd,pos/.initial=0.35,height/.initial=0}

\usepackage{lstparams}
\usepackage{lstcoq}

\usepackage[capitalize, nameinlink]{cleveref}

\usepackage{siunitx}
\newcommand\cdc[1]{\lstinline[language=Coq,breakatwhitespace]{#1}}
\newcommand\zcdc[1]{\let\par\endgraf\cdc{#1}}

\newcommand\hstocoq{\texttt{hs-to-coq}\xspace}
\newcommand{\LiquidHaskell}{\textsc{Liquid Haskell}\xspace}
\newcommand\irisc{Iris$^{\$}$\xspace}

\newcommand\etc{\textit{etc.}}
\newcommand\ie{\textit{i.e.,\ }}
\newcommand\eg{\textit{e.g.,\ }}

\newcommand\method{reverse physicist's method}

\newcommand\MethodTitle{Reverse Physicist's Method}
\newcommand\potential{potential}

\newboolean{extended}
\newboolean{review}

\usepackage{draft} 
\newnote{scw}{blue}       
\newnote{ly}{violet}      
\newnote{lx}{olive}      
\newnote{lai}{cyan}    

\draftfalse
\extendedtrue
\reviewfalse

\setcopyright{rightsretained}
\acmJournal{PACMPL}
\acmYear{2024}
\acmVolume{8}
\acmNumber{ICFP}
\acmArticle{237}
\acmMonth{8}
\acmDOI{10.1145/3674626}

\newcommand\evalcolor{\color{blue}}
\newcommand\demcolor{\color{purple}}
\newcommand\deneval[1]{{\evalcolor\llbracket} #1 {\evalcolor\rrbracket}_{\evalcolor\mathrm{eval}}}
\newcommand\dendem[1]{{\demcolor\llbracket} #1 {\demcolor\rrbracket}_{\demcolor\mathrm{dem}}}
\newcommand\denevalT[1]{{\evalcolor\llbracket} #1 {\evalcolor\rrbracket}_{\evalcolor\mathrm{eval}}}
\newcommand\denapproxT[1]{{\demcolor\llbracket} #1 {\demcolor\rrbracket}_{\demcolor\mathrm{approx}}}
\newcommand\denapproxTT[2]{{\demcolor\llbracket} #1 {\demcolor\rrbracket}^{#2}_{\demcolor\mathrm{approx}}}
\newcommand\dencv[1]{{\demcolor\llbracket} #1 {\demcolor\rrbracket}_{\demcolor\mathrm{cv}}}
\newcommand\lubplus{\sqcupplus}
\newcommand*\alt{\;|\;}
\newcommand*\tlet[3]{\mathrm{let}\,#1=#2\,\mathrm{in}\,#3}

\begin{document}
\title{Story of Your Lazy Function's Life}
\subtitle{A Bidirectional Demand Semantics for Mechanized Cost Analysis of Lazy Programs}

\author{Li-yao Xia}
\email{lyxia@poisson.chat}
\orcid{0000-0003-2673-4400}
\affiliation{%
  \institution{unaffiliated}
  \country{France}
}


\author{Laura Israel}
\email{laisrael@pdx.edu}
\orcid{0009-0003-5008-1785}
\affiliation{%
  \institution{Portland State University}
  \city{Portland}
  \state{OR}
  \country{USA}
}

\author{Maite Kramarz}
\email{maite.kramarz@mail.utoronto.ca}
\orcid{0009-0006-0623-2293}
\affiliation{%
  \institution{University of Toronto}
  \city{Toronto}
  \state{Ontario}
  \country{Canada}
}

\author{Nicholas Coltharp}
\email{coltharp@pdx.edu}
\orcid{0000-0002-4832-7016}
\affiliation{%
  \institution{Portland State University}
  \city{Portland}
  \state{OR}
  \country{USA}
}

\author{Koen Claessen}
\email{koen@chalmers.se}
\orcid{0000-0002-8113-4478}
\affiliation{%
  \institution{Chalmers University of Technology}
  \city{Gothenburg}
  \country{Sweden}
}

\author{Stephanie Weirich}
\email{sweirich@cis.upenn.edu}
\orcid{0000-0002-6756-9168}
\affiliation{%
  \institution{University of Pennsylvania}
  \city{Philadelphia}
  \state{PA}
  \country{USA}
}

\author{Yao Li}
\email{liyao@pdx.edu}
\orcid{0000-0001-8720-883X}
\affiliation{%
  \institution{Portland State University}
  \city{Portland}
  \state{OR}
  \country{USA}
}

\begin{abstract}
  Lazy evaluation is a powerful tool that enables better compositionality and
  potentially better performance in functional programming, but it is
  challenging to analyze its computation cost. Existing works either require
  manually annotating sharing, or rely on separation logic to reason about heaps
  of mutable cells. In this paper, we propose a bidirectional demand semantics
  that allows for extrinsic reasoning about the computation cost of lazy programs without
  relying on special program logics. To show the effectiveness of our approach,
  we apply the demand semantics to a variety of case studies including insertion
  sort, selection sort, Okasaki's banker's queue, and the implicit queue. We
  formally prove that the banker's queue and the implicit queue are both
  amortized and persistent using the Rocq Prover (formerly known as Coq). We
  also propose the reverse physicist's method, a novel variant of the classical
  physicist's method, which enables mechanized, modular and compositional
  reasoning about amortization and persistence with the demand semantics.
\end{abstract}



\maketitle

\bibliographystyle{ACM-Reference-Format}
\citestyle{acmauthoryear}   

\section{Introduction}\label{sec:intro}

The power of laziness is great, but formal reasoning about its costs is
notoriously elusive. After all, lazy evaluation is stateful and produces
interleaved computation, with the cost of functions depending on future demand.
We believe that mechanized reasoning can help ensure the rigor of the analysis
needed to understand these costs. To realize this vision, we present a
shallow-embedding-based model of cost analysis. Our approach allows us to
mechanically and extrinsically reason about the computational cost of lazy
functional programs and lazy, amortized, and persistent functional data
structures.

Our solution is based on a \emph{bidirectional demand semantics}. The semantics
was first described by \citet{bjerner1989} in an untyped setting. We adapt and
expand it to a typed and total semantics. Given a lazy function $f : A \to B$,
we can use the bidirectional demand semantics to systematically derive a demand
function $f^D : A \to B^D \to \mathbb{N} \times A^D$, where $A^D$ represents the
demand on type $A$ and $\mathbb{N}$ is the computation cost. That is, given the
input ($A$) to function $f$ and the demand on its output ($B^D$), the
\emph{demand function} $f^D$ calculates the minimal demand on the input ($A^D$),
as well as the computation cost required to obtain the demanded
output~($\mathbb{N}$). In a sense, the demand function tells the ``story of a
lazy function's life,'' including what happens when it is subjected to future
demands. We can calculate such an input demand because, in a deterministic
language, given any valid input and output demand, there exists \emph{exactly
  one} minimal input demand. The use of input/output demand and demand functions
distinguishes our work from alternative approaches based on heaps of mutable
cells, such as \citet{iris-credit} and \citet{iris-thunk}, which rely on
separation logic for reasoning.

To demonstrate the effectiveness of our demand semantics, we have developed a
wide variety of case studies. We use our model to formally prove the computation
cost of lazy insertion sort, lazy selection sort,
\citeauthor{{purely-functional}}'s \emph{banker's queue} and \emph{implicit
  queue}~\citep{purely-functional}. For the banker's queue and the implicit
queue, we also show that these data structures are both amortized and
persistent.

To reason about amortization and persistence in a modular way, we propose the
\emph{\method}, a novel variant of the classical physicist's method for
amortized computational complexity analysis in strict
semantics~\cite{amortized}. Similar to the classical physicist's method, the
\method{} makes use of a \emph{\potential{}} function, which we apply to
approximations of datatypes to describe their \emph{accumulated} \potential{}.
All proofs are mechanized using the Rocq Prover~(formerly known as the Coq
theorem prover). All Rocq Prover definitions and proofs can be found in our
artifact, which is publicly available~\citep{demand-artifact}.

In summary, we make the following contributions:
\begin{itemize}
  \item We propose a bidirectional demand semantics for lazy functional programs
        based on \citet{bjerner1989}~(\Cref{sec:demand-semantics}). Our demand
        semantics is typed and total, allowing it to be formalized in proof
        assistants such as the Rocq Prover.
  \item We formally prove that the bidirectional demand semantics is equivalent
        to the natural semantics of laziness~\citep{launchbury1993} in the Rocq
        Prover, by showing its equivalence to another semantics that is
        equivalent to natural semantics, namely clairvoyant
        semantics~\citep{clairvoyant, forking-paths}~(\Cref{subsec:equiv}).
  \item We show how the demand semantics can be used to systematically derive
        demand functions for a realistic programming language by proving
        computation cost theorems for insertion sort and selection
        sort~(\cref{sec:case-study-1}).
  \item We propose the \emph{\method{}}, a novel method for analyzing amortized
        computation cost and persistence for lazy functional data structures
        based on demand semantics~(\cref{sec:physicist}).
  \item We present a \emph{mechanized proof} in the Rocq Prover which shows that
        \citeauthor{purely-functional}'s banker's queue and implicit queue are
        amortized and persistent using the \method. Our mechanized proof does
        not rely on trusting the demand
        functions~(\cref{subsec:banker}--\ref{subsec:implicit}). 
\end{itemize}

In addition, we provide a sketch of our method with motivating examples in
\Cref{sec:sketch}. We discuss related work in \Cref{sec:related-work} and future
work in \Cref{sec:conclusion}.

\section{Motivating Examples}\label{sec:sketch}

\subsection{A Demand Semantics}

To introduce and motivate our method of reasoning about lazy programs, we will
consider \cdc{insertion\_sort} as an example. This algorithm is known to run in
$O(n^2)$ time for a list of length $n$ under eager evaluation. In contrast, a
lazy implementation only computes each successive sorted element when a result
is demanded, potentially causing asymptotically lower time costs if only part of
the list is needed.

To model the improvements to the computation cost of a lazy
\cdc{insertion\_sort}, we first compose the \cdc{take} and \cdc{insertion_sort}
functions (\cref{fig:take-insertion-sort}). We implement these functions in
Gallina, the underlying specification language of the Rocq Prover. Even though
Gallina is not a lazy language, we can imagine a ``translator'' that converts
these functions from a lazy functional language to Gallina~(\eg~\hstocoq for
Haskell~\citep{containers}).

\begin{figure}[t]
\begin{lstlisting}[language=Coq,numbers=left,xleftmargin=4.0ex]
Fixpoint insert (x : nat) (xs : list nat) : list nat :=
  match xs with 
  | [] => x :: []
  | y :: ys => if y <=? x then
                 let zs := insert x ys in
                 y :: zs
               else x :: y :: ys
  end.

Fixpoint insertion_sort (xs : list nat) : list nat :=
  match xs with
  | nil => nil
  | y :: ys =>
      let zs := insertion_sort ys in
      insert y zs
  end.
\end{lstlisting}
\caption{The Gallina implementation of \cdc{insert} and \cdc{insertion_sort} in
  ANF~(A-normal form)~\citep{anf}. For simplicity, we define these functions on
  lists of natural numbers~(\cdc{nat}). The infix operator \cdc{<=?} shown in
  \cdc{insert} is Gallina's ``less than or equal'' operator on natural numbers,
  which returns a boolean.}\label{fig:take-insertion-sort}
\end{figure}

\paragraph{Representing demand}

To model laziness, we first need a notion of input and output demand. We use the
\emph{approximation data types} proposed by \citet{forking-paths} to represent
both \emph{approximations} and \emph{demands}. An approximation is a partial
value with a placeholder for unevaluated thunks. For example, the finite list
data structure has the approximation data type \cdc{listA}:
\begin{lstlisting}[language=Coq]
Inductive listA (a : Type) : Type :=
  NilA | ConsA (x : T a) (xs : T (listA a)).
\end{lstlisting}
The \cdc{T} data type is a sum type that can be either an unevaluated thunk
\cdc{Undefined} (also denoted as $\bot$) or an evaluated value \cdc{Thunk a}.
Approximation data types can be used to specify demands---\eg~a \cdc{listA
  nat} of \cdc{ConsA (Thunk 0) Undefined} represents a demand on a \cdc{list
  nat} such that only the first item in the list must be evaluated \emph{and} it
evaluates to 0.

\paragraph{Definedness ordering}

Before we can state any theorems, we need a definedness order between ordinary
data types and approximation data types, as well as between different
approximation data types. For example, consider the following two approximation
data types:
\begin{lstlisting}[language=Coq]
Definition lA1 := ConsA (Thunk 0) Undefined.
Definition lA2 := ConsA (Thunk 0) (ConsA (Thunk 1) Undefined).
\end{lstlisting}
We say that \cdc{lA1} is \emph{less defined} than \cdc{lA2} because \cdc{lA2} is
defined on the first and second elements of the list, while \cdc{lA1} is
defined only on the first. We also say that both \cdc{lA1} and \cdc{lA2} are
\emph{approximations of} \cdc{[0; 1; 2]}, but only \cdc{lA1} is an approximation
of \cdc{[0; 2]}. We defer formal definitions of these definedness orders to
\cref{subsec:lazy-semantics}.

\paragraph{Demand functions}

As we saw in the introduction, each lazy function has a corresponding \emph{demand function}. Given an
\emph{output demand}, the demand function computes the \emph{minimal input
  demand} required for the function to satisfy the output demand, as well as the
time cost incurred. We show the demand functions of \cdc{insert} and
\cdc{insertion_sort} in \cref{fig:insertD}. As a convention, we suffix a
function's name with capital letter \cdc{D} to indicate that it is a demand
function.

\begin{figure}[t]
\begin{lstlisting}[language=Coq,numbers=left,xleftmargin=4.0ex]
Fixpoint insertD (x:nat) (xs: list nat)
  (outD : listA nat) : Tick (T (listA nat)) :=
  tick >>
  match xs, outD with 
  | [], ConsA zD zsD =>
      ret (Thunk NilA)
  | y :: ys, ConsA zD zsD => 
     if y <=? x then 
       let+ ysD := thunkD (insertD x ys) zsD in
       ret (Thunk (ConsA (Thunk y) ysD))
     else 
       ret zsD
  | _ , _ => bottom (* absurdity case *)
  end.

Fixpoint insertion_sortD (xs: list nat)  (outD : listA nat) :
  Tick (T (listA nat)) :=
  tick >>
  match xs with
  | [] => ret (Thunk NilA)
  | y :: ys =>
      let zs := insertion_sort ys in
      let+ zsD := insertD y zs outD in
      let+ ysD := thunkD (insertion_sortD ys) zsD in
      ret (Thunk (ConsA (Thunk y) ysD))
  end.
\end{lstlisting}
\caption{The demand functions of \cdc{insert} and \cdc{insertion_sort}.}\label{fig:insertD}
\end{figure}

For a concrete example, we first look at \cdc{insertD}, the demand function of \cdc{insert}. In addition
to the arguments of \cdc{insert}~(line~1), the demand function \cdc{insertD}
takes an extra argument \cdc{(outD : listA nat)}, which represents the output
demand~(line~2). The function returns the minimal input demand \cdc{T (listA
  nat)} and wraps it in a \cdc{Tick} data type~(line~2). \cdc{Tick} is a monad defined as \cdc{Tick a = nat
  * a}, where \cdc{nat} is the type of natural numbers representing the time cost of a wrapped function. It is essentially a writer monad specialized to \cdc{nat}. \ly{Talk
  about why we don't return nat as well.}

The \cdc{tick} operation increments the time cost by one~(line~3). We count the number
of function calls by invoking \cdc{tick} at the beginning of every function.
The function then matches on its input as well as the output demand~(line~4).

If
\cdc{xs} is an empty list, we know from the definition of \cdc{insert} that the
output demand must be an approximation of a \cdc{x :: nil}, so we also match
\cdc{outD} to ensure that it has the form \cdc{ConsA zD zsD}~(line~5). Based on the implementation of \cdc{insert}, we also must evaluate the pattern matching to determine that
the input is empty, so we return the minimal input demand as
\cdc{NilA}~(line~6).

If \cdc{xs} is not an empty list, we know from \cdc{insert} that the
output must be an approximation of a non-empty list, so we match \cdc{outD} to
the form \cdc{ConsA zD zsD}~(line~7). Like in \cdc{insert}, we then proceed to
check whether \cdc{y <=? x}~(line~8). If \cdc{y <=? x}, we make a recursive call to
\cdc{insertD} to get the input demand of the recursive call to
\cdc{insert}~(lines~9--10). Otherwise, we return \cdc{zsD}~(line~12). The
\cdc{let+} notation on line~9 is a custom notation for bind in the \cdc{Tick} monad.
The \cdc{thunkD} combinator on line~9 is a function of type \cdc{(A -> B)
  -> T A -> T B}, \ie~it applies a function to data wrapped in a Thunk.

The demand function \cdc{insertD} has a similar structure to \cdc{insert}.
This is not a coincidence. We will show that, given a pure function, we can
systematically derive its demand function in \cref{sec:demand-semantics}.

\paragraph{Bidirectionality}

\begin{figure}[t]
\includegraphics[page=1,width=0.7\textwidth]{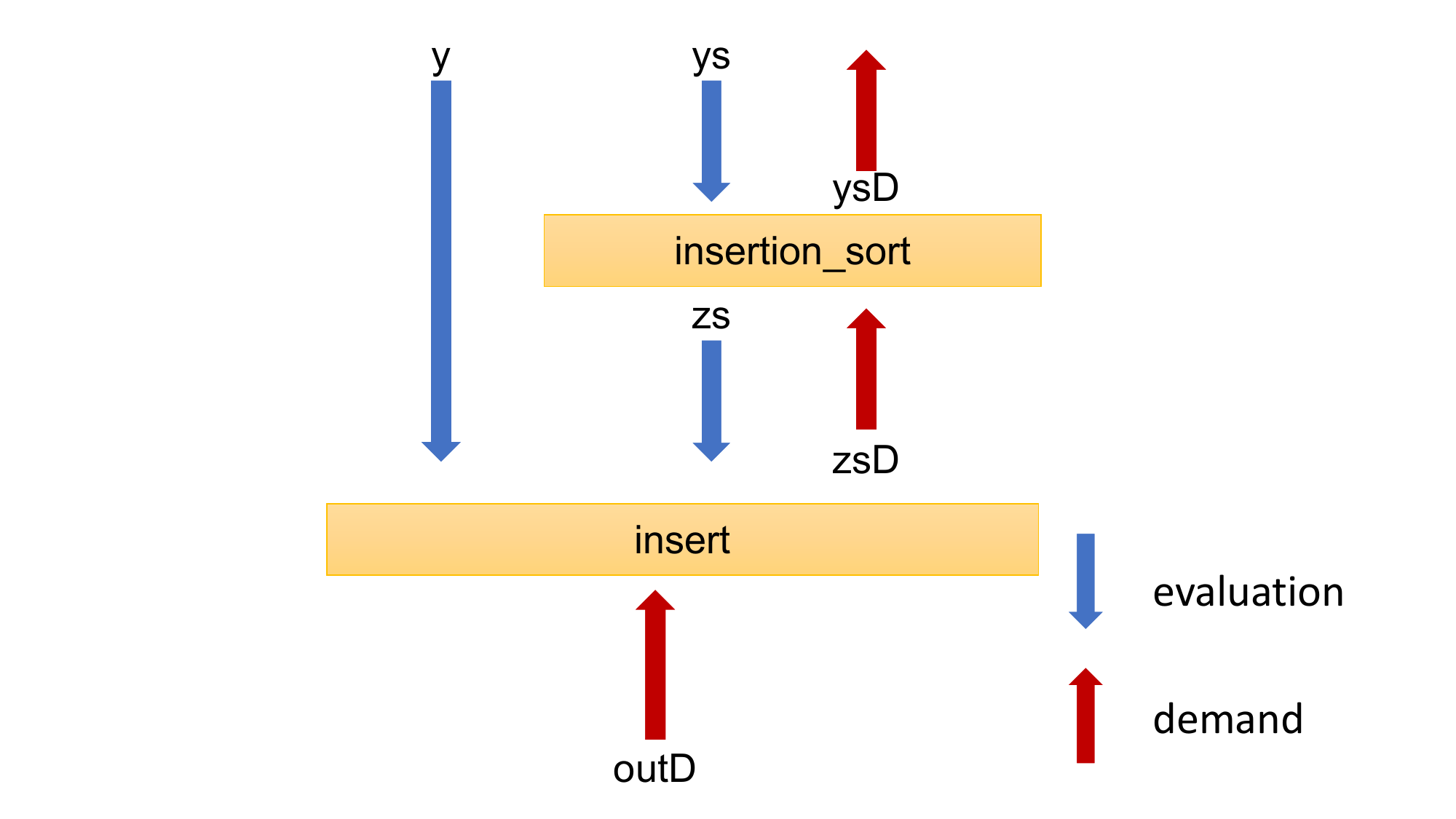}
\caption{An illustration of how the input demand is computed in
  \cdc{insertion_sortD}.}\label{fig:insertion_sortD_illustration}
\end{figure}

\Cref{fig:insertD} shows \cdc{insertion_sortD}, the demand function of
\cdc{insertion_sort}, which has a more advanced implementation than
\cdc{insertD}. The first case~(line~20) is straightforward, but the second
case~(line~21) is more complex. This is because \cdc{insertion_sort} first calls
itself recursively, and then applies \cdc{insert} to the result of the recursive
call (lines~14--15 in \cref{fig:take-insertion-sort}). In
\cref{fig:insertion_sortD_illustration}, we illustrate how we compute the input
demand for \cdc{insertion_sortD} via a demand dependency graph of
\cdc{insertion_sort}. We start with the input \cdc{y} and \cdc{ys}, as well as
the output demand \cdc{outD}. To calculate the input demand of a
\cdc{insertion_sort}, we need the input demand of the recursive call \cdc{ysD}.
However, \cdc{ysD} relies on the input \cdc{ys} and the output demand of the
recursive call \cdc{zsD}, which in turn relies on the input \cdc{y}, the output
demand \cdc{outD}, and the input \cdc{zs} which is the result of evaluating
\cdc{insertion_sort}.

\cref{fig:insertion_sortD_illustration} shows that the demand function needs to run computation in both
directions: an \emph{evaluation} direction that computes output from input~(blue
arrows in the figure), and a \emph{demand} direction that computes the input
demand from pure input and output demand (red arrows in the figure). It
also reveals how we should define \cdc{insertion_sortD}. Starting from the top, we first call
\cdc{insertion_sort}~(line~22, \cref{fig:insertD}), then we call
\cdc{insertD}~(line~23), and finally \cdc{insertion_sortD}~(line~24).

\paragraph{Specification and proof sketches}

From here, we can prove properties about these demand functions. For
example, we can prove the following theorems for \cdc{insertion_sortD}:
\begin{lstlisting}[language=Coq]
Theorem insertion_sortD_approx (xs : list nat) (outD : listA nat)
  : outD `is_approx` insertion_sort xs ->
    Tick.val (insertion_sortD xs outD) `is_approx` xs.

Theorem insertion_sortD_cost (xs : list nat) (outD : listA nat) :
  Tick.cost (insertion_sortD xs outD) <= (sizeX' 1 outD + 1) * (length xs + 1).
\end{lstlisting}

The theorem \cdc{insertion_sortD_approx} describes the \emph{functional
  correctness} of \cdc{insertion_sortD}. It states that, if the given output
demand is an approximation of the output, then the input demand is an
approximation of the original input.

The theorem \cdc{insertion_sortD_cost} describes the \emph{time cost} of
\cdc{insertion_sortD}. It states that the cost of \cdc{insertion_sort} is
bounded by $(\text{max}(1, |\texttt{outD}|) + 1) \times (|\texttt{xs}| + 1)$. In
other words, for each element of the sorted list we compute, we will have to
linearly search through the input list once (plus some constant overhead). This
is an overestimation---the input list will shrink after each recursive call and
we don't have to go through the entire list when inserting a new elements---but
this still provides a useful upper bound.

As programmers, we may also be interested in the demands exerted by functions
that call the functions we analyze. Using \cdc{take : nat -> list A ->
  list A} as an example: the function call \cdc{take n xs} takes the first
\cdc{n} cells from list \cdc{xs}. We can prove that the demand of \cdc{takeD n
  xs outD} has a size bounded by the length of \cdc{outD}. Using this lemma, we
can then describe the cost of running a function composing \cdc{take} with
\cdc{insertion_sort}, and then prove that the composed function satisfies that
cost. This new cost is linear with respect to \cdc{n}, the first parameter of
\cdc{take}. When \cdc{n} is a constant, the cost is an asymptotic improvement
over the eager version of \cdc{insertion_sort}. This formalizes a common pattern
in lazy programming---computing only the necessary parts of an expensive
function call to reduce costs.

This example demonstrates that our approach is compositional. Even though a
functional call under lazy evaluation is not local (as a future demand may cause
the function to run further), we can specify each function individually using
demand semantics. If we wish to compose these functions with others and reason
about the cost of their composition, we can do so by proving theorems for
external functions individually and composing their specifications. For example,
we may want to compose \cdc{take} with a different sorting algorithm,
\eg~\cdc{selection_sort}. We demonstrate how we use demand semantics to reason
about all these interactions in \cref{sec:case-study-1}.

\subsection{Amortized Analysis for Persistent Data Structures}

Besides the usual benefits of lazy evaluation such as
compositionality~\citep{fp-matters}, lazy evaluation also enables
the combination of \emph{persistence} (old values can be reused) and
\emph{amortization} (we average the cost of operations over multiple
calls)~\citep{purely-functional}. For example, a call to \cdc{pop} on a
banker's queue may have a high initial cost to evaluate many thunks, so that
subsequent calls to \cdc{pop} again on the same old queue will be cheap.
That is not possible with eager evaluation, where all repeated calls on the
same value have the same cost.

In the Rocq Prover, we formally verified amortized cost bounds for both 
the banker's queue and the implicit queue under persistence. Our final theorems 
show that the cost of executing a program trace of either the banker's queue or 
the implicit queue is always linear with respect to the number of operations.

Proving these theorems directly is challenging. Instead of reckoning with
arbitrary program traces for each queue, we develop a modular framework that
allows us to reason about the cost of both queues' functions based on their
respective demand functions. The framework is based on the \emph{\method}, a
novel variant of the classical physicist's method that we propose in this paper.
We provide more details about the verification of the banker's queue and
implicit queue in \cref{sec:physicist}.

\section{Bidirectional Demand Semantics}\label{sec:demand-semantics}

In this section, we show the definition of the demand
semantics~(\cref{subsec:lazy-semantics}) and its
properties~(\cref{subsec:properties}). To show that the demand semantics
correctly models laziness, we show its equivalence with natural
semantics~\citep{launchbury1993} by showing its equivalence to clairvoyant
semantics~\citep{clairvoyant, forking-paths}~(\cref{subsec:equiv}). All the
lemmas and theorems shown in this section have been formally proven in the Rocq
Prover.

\subsection{Lazy Semantics}\label{subsec:lazy-semantics}

\paragraph{Syntax}
We consider a pure, total, first-order calculus with explicit thunks. Evaluation
is eager by default. This calculus can be viewed either as a subset of ML with a
type for memoized thunks~(\eg~\texttt{lazy} in OCaml, denoted \texttt{T} here),
similar to the language used in \citet{purely-functional}, or as an intermediate
representation which lazy languages such as Haskell can be translated into.\ly{A
  little bit more on the difficulty of translating from Haskell to this?} Making
explicit the constructs ($\mathrm{lazy}$ and $\mathrm{force}$) to manipulate
thunks makes our semantics rather simple.

We show the syntax of the language in \cref{fig:syntax} and the typing rules in
\cref{fig:typing}. The language includes types such as booleans, lists,
products, and a thunk type \cdc{T} that is a sum type of unevaluated thunk and
evaluated value.\footnote{This is the same datatype \cdc{T} as shown in
  \cref{sec:sketch}.} The language is first-order: higher-order functions are
not allowed, as evidenced by the lack of function type, but it is equipped with
a primitive $\mathrm{foldr}$ for defining recursions. Most of the language's
operators are standard. The $\mathrm{tick}$ operation increases the count of the
current computation cost by 1. The $\mathrm{lazy}$ and $\mathrm{force}$
operations are the opposite of each other: $\mathrm{lazy}$ creates a new thunk
\cdc{T} with suspended computation and $\mathrm{force}$ triggers the suspended
computation in a thunk~(and does nothing if it's a value).

\begin{figure}
\begin{align*}
  A,B \;::=\;& \mathrm{bool} \alt \mathrm{list}\,A \alt \mathrm{T}\,A \alt A \times B\\
  M,N \;::=\;& x \alt \tlet{x}{M}{N} \alt \mathrm{tick}\,M \alt \mathrm{lazy}\,M \alt \mathrm{force}\,M \\
    \alt & \mathrm{cons}\,M\,N \alt \mathrm{nil} \alt \mathrm{foldr}\,(\lambda\,x\,y.\,M_1)\,M_2\,M_3 \\
    \alt & \mathrm{pair}\,M\,N \alt \mathrm{fst}\,M \alt \mathrm{snd}\,M \alt \mathrm{true} \alt \mathrm{false} \alt \mathrm{if}\,M_1\,M_2\,M_3
\end{align*}
  \caption{Syntax}
  \label{fig:syntax}
\end{figure}

\begin{figure}
\begin{mathpar}
  \inferrule[Var]
    {\Gamma(x) = A}
    {\Gamma \vdash x : A}
  \quad
  \inferrule[Let]
    {\Gamma \vdash M : A \\ \Gamma, x:A \vdash N : B}
    {\Gamma \vdash \tlet{x}{M}{N} : B}
  \quad
  \inferrule[Tick]
    {\Gamma \vdash M : A}
    {\Gamma \vdash \mathrm{tick}\,M : A}
  \\
  \inferrule[Lazy]
    {\Gamma \vdash M : A}
    {\Gamma \vdash \mathrm{lazy}\,M : T A}
  \quad
  \inferrule[Force]
    {\Gamma \vdash M : T\,A}
    {\Gamma \vdash \mathrm{force}\,M : A}
  \\
  \inferrule[Cons]
    {\Gamma \vdash M : T\,A \\ \Gamma \vdash N : T\,(\mathrm{list}\,A)}
    {\Gamma \vdash \mathrm{cons}\,M\,N : \mathrm{list}\,A}
  \\
  \inferrule[Foldr]
    {\Gamma, x : T\,A, y : T\,B \vdash M_1 : B \\ \Gamma \vdash M_2 : B \\ \Gamma \vdash M_3 : \mathrm{list}\,A}
    {\Gamma \vdash \mathrm{foldr}\,(\lambda\,x\,y.\,M_1)\,M_2\,M_3 : B}
\end{mathpar}
  \caption{Typing rules.}\label{fig:typing}
\end{figure}

\begin{figure}[t]
\begin{align*}
  \denevalT{A} & : \mathrm{Set} \\
  \denevalT{\mathrm{bool}} & = \{0,1\} \\
  \denevalT{\mathrm{T}\,A} & = \denevalT{A} \\
  \denevalT{\mathrm{list}\,A} & = \{ \mathrm{nil} \} \uplus \left\{ \mathrm{cons}\,a\,b \mid a \in \denevalT{A}, b \in \denevalT{\mathrm{list}\,A} \right\}
\end{align*}
                               
\begin{align*}
  \denapproxT{A} & : \mathrm{Set} \\
  \denapproxT{\mathrm{bool}} & = \{0,1\} \\
  \denapproxT{\mathrm{T}\,A} & = \mathrm{T}\,\denapproxT{A} \overset{\mathrm{def}}{=} \{ \bot \} \uplus \left\{ \mathrm{thunk}\,a \mid a \in \denapproxT{A} \right\} \\
  \denapproxT{\mathrm{list}\,A} & = \{ \mathrm{nil} \} \uplus \left\{ \mathrm{cons}\,a\,b \mid a \in \mathrm{T}\,\denapproxT{A}, b \in \mathrm{T}\,\denapproxT{\mathrm{list}\,A} \right\}
\end{align*}
\caption{Sets of total values $\denevalT{A}$ and sets of approximations
  $\denapproxT{A}$}\label{fig:approx}
\end{figure}

\begin{figure}[t]
\begin{align*}
  \deneval{\Gamma \vdash M : A} & : \denevalT{\Gamma} \to \denevalT{A} \\
  \deneval{x}(g) & = g(x) \\
  \deneval{\mathrm{force}\,M}(g) & = \deneval{M}(g) \\
  \deneval{\mathrm{lazy}\,M}(g) & = \deneval{M}(g) \\
  \deneval{\tlet{x}{M}{N}}(g) & = \deneval{N}(\{g, x \mapsto\deneval{M}(g)\}) \\
  \deneval{\mathrm{cons}\,M\,N}(g) & = \mathbf{cons}\,\deneval{M}(g)\,\deneval{N}(g) \\
  \deneval{\mathrm{nil}}(g) & = \mathbf{nil} \\
  \deneval{\mathrm{foldr}\,(\lambda x y. M_1)\,M_2\,N}(g) & = \mathbf{foldr}_{\evalcolor\mathrm{eval}}(g,M_1,M_2,\deneval{N}(g)) \\
  \mathbf{foldr}_{\evalcolor\mathrm{eval}}(g,M_1,M_2,\mathbf{nil}) & = \deneval{M_2}(g) \\
  \mathbf{foldr}_{\evalcolor\mathrm{eval}}(g,M_1,M_2,(\mathbf{cons}\,a_1\,a_2)) & = \deneval{M_1}(\{g, x \mapsto a_1, y \mapsto \mathbf{foldr}_{\evalcolor\mathrm{eval}}(g,M_1,M_2,a_2)\})
\end{align*}
  \caption{Forward evaluation.}
\end{figure}

\begin{figure}[t]
\begin{align*}
  (c_M, d_M) \lubplus (c_N, d_N) & = (c_M + c_N, d_M \sqcup d_N) \\
  \dendem{\Gamma \vdash M : A} & : \denevalT{\Gamma} \times \denapproxT{A} \rightharpoonup \mathbb{N} \times \denapproxT{\Gamma} \\
  \dendem{x}(g, a) & = (0, \{ x \mapsto a \}) \\
  \dendem{\mathrm{tick}\,M}(g, a) & = (1 + c, d) \quad\text{ where }(c, d) = \dendem{M}(g, a) \\
  \dendem{\mathrm{force}\,M}(g, a) & = \dendem{M}(g, \mathbf{thunk}\,a) \\
  \dendem{\mathrm{lazy}\,M}(g, \bot) & = (0, \bot_g) \\
  \dendem{\mathrm{lazy}\,M}(g, \mathbf{thunk}\,a) & = \dendem{M}(g, a) \\
  \dendem{\tlet{x}{M}{N}}(g, a) & = (c_N + c_M, d_N \sqcup d_M) \\[-2mm]
    \intertext{\raggedleft
      $\begin{aligned}
        \text{where } & (c_N, \{ d_N, x \mapsto b \}) = \dendem{N}(\{g,x \mapsto \deneval{M}(g)\} , a) \\
        \text{and } & (c_M, d_M) = \dendem{M}(g, b)
      \end{aligned}$}
  \dendem{\mathrm{cons}\,M\,N}(g, \mathbf{cons}\,a\,b) & = \dendem{M}(g, a) \lubplus \dendem{N}(g, b) \\
  \dendem{\mathrm{nil}}(g, \mathbf{nil}) & = (0, \bot) \\
  \dendem{\mathrm{foldr}\,(\lambda x y. M_1)\,M_2\,N}(g, d) & = (c, g') \lubplus \dendem{N}(g, n') \\[-2mm]
    \intertext{\raggedleft where $(c, g', n') = \mathbf{foldr}_{\demcolor\mathrm{dem}}(g,M_1,M_2,\mathbf{thunk}\,(\deneval{N}(g)),\mathbf{thunk}\,d)$}
\end{align*}
\caption{Demand semantics: backward evaluation.}\label{fig:demand-semantics}
\end{figure}
\begin{figure}[t]
\begin{align*}
  \mathbf{foldr}_{\demcolor\mathrm{dem}}(g,M_1,M_2,n,\bot) & = (0, \bot_g, \bot) \\
  \mathbf{foldr}_{\demcolor\mathrm{dem}}(g,M_1,M_2,\mathbf{thunk}\,\mathbf{nil},\mathbf{thunk}\,d) & = \dendem{M_2}(g,d) \\
  \mathbf{foldr}_{\demcolor\mathrm{dem}}(g,M_1,M_2,\mathbf{thunk}\,(\mathbf{cons}\,a_1\,a_2),\mathbf{thunk}\,d) & = (c_1 + c_2, g_1 \sqcup g_2, \mathbf{thunk}\,(\mathbf{cons}\,a_1'\,a_2'))\\[-2mm]
    \intertext{\raggedleft
      $\begin{aligned}
        \text{where } & (c_1,\{g_1,x \mapsto a_1',y\mapsto b_2'\}) = \dendem{M_1}(\{g,x\mapsto a_1,y\mapsto \mathbf{foldr}_{\evalcolor\mathrm{eval}}(g,M_1,M_2, a_2)\}, d) \\
        \text{and } & (c_2, g_2, a_2') = \mathbf{foldr}_{\demcolor\mathrm{dem}}(g,M_1,M_2,a_2,b_2')
      \end{aligned}$}
\end{align*}
  \caption{Definition of $\mathbf{foldr}_{\demcolor\mathrm{dem}}$}\label{fig:foldr-dem}
\end{figure}

\paragraph{Semantics}
Given a lazy function, we can automatically translate it to a demand function.
We present such a translation as a denotational semantics. The semantics is
compositional: the denotation of a term is depends only on the denotation of its
immediate subterms.\ly{I have trouble understanding this sentence. Can you
  rephrase?} \lai{Me too, I'm having a hard time parsing it.} It can be
viewed as a translation from our calculus to a metalanguage able to express
our semantics, which mainly consists of pattern-matching and
recursion on the list approximation type.
The denotational semantics consists of two
denotation functions on well-typed terms $\Gamma \vdash M : A$, a pure forward
interpretation $\deneval{M}$ and a demand interpretation $\dendem{M}$.

\paragraph{Forward evaluation}
The ``forward evaluation'' $\deneval{M} : \denevalT{\Gamma} \to \denevalT{A}$ is
the natural functional interpretation. Our calculus is total, so all terms have
a value. Thunks can always be evaluated, so the interpretation of
a lifted type $\denevalT{T\,A}$ is the same as the unlifted $\denevalT{A}$.
The demand semantics~\citep{bjerner1989} is defined by ``backwards evaluation'':
$\dendem{M} : \denevalT{\Gamma} \times \denapproxT{A} \rightharpoonup \mathbb{N} \times \denapproxT{\Gamma}$.

\paragraph{Lattice of approximations}
Intuitively, lazy evaluation is driven by demand: the evaluation of a term
depends on how much of its result will be needed. Let us first describe the
representation and structure of demands as \emph{approximations}. The set of
approximations $\denapproxT{A}$ consists of values with the same shape as in
$\denevalT{A}$, possibly with some subterms replaced with a special value $\bot$
representing an unneeded thunk. $\denapproxT{A}$ is defined formally in
\Cref{fig:approx}. Approximations are ordered by definedness. This partial
order, denoted $a \le b$, is defined inductively in \Cref{fig:definedness}:
$\bot \le a$ for all $a$. The $\le$ relation is reflexive, and all constructors
($\mathbf{cons}$ and $\mathbf{thunk}$ in our calculus) are monotone. By
contrast, we say that elements of $\denevalT{A}$ are \emph{total values}.

\begin{figure}
  \begin{mathpar}
    \inferrule
      { }
      {a \le a}
      {\hbox{reflexivity}}
      \quad
    \inferrule
      { }
      {\bot \le a}
      {\bot\hbox{-least}}
      \quad
    \inferrule
      {a \le b}
      {\mathbf{thunk}\,a \le \mathbf{thunk}\,b}
      {\mathbf{thunk}}
      \quad
    \inferrule
      {a \le c \\ b \le d}
      {\mathbf{cons}\,a\,b \le \mathbf{cons}\,c\,d}
      {\mathbf{cons}}
  \end{mathpar}
  \caption{Definedness order $a \le b$}
  \label{fig:definedness}
\end{figure}

\begin{figure}
\begin{minipage}{0.6\textwidth}
  \begin{mathpar}
    \inferrule
      { }
      {\bot \prec a}
      {\bot\hbox{-least}}
      \quad
    \inferrule
      {c \in \{ \mathrm{nil}, \mathrm{true}, \mathrm{false} \}}
      {c \prec c}
      {\hbox{$\prec$-$c$}}
      \\
    \inferrule
      {a \prec b}
      {\mathbf{thunk}\,a \prec b}
      {\hbox{$\prec$-$\mathbf{thunk}$}}
      \quad
    \inferrule
      {a \prec c \\ b \prec d}
      {\mathbf{cons}\,a\,b \prec \mathbf{cons}\,c\,d}
      {\hbox{$\prec$-$\mathbf{cons}$}}
  \end{mathpar}
  \caption{Approximation relation $a \prec b$}
  \label{fig:definedness}
\end{minipage}
\begin{minipage}{0.35\textwidth}
  \begin{align*}
    \bot_{-} & : \denevalT{A} \to \denapproxT{A} \\
    \bot_x & = \bot \quad\text{if } \exists B,\, A = T B \\
    \bot_{\mathbf{cons}\,a\,b} & = \mathbf{cons}\,\bot\,\bot \\
    \bot_{\mathbf{nil}} & = \mathbf{nil} \\
    \bot_{\{x_i \mapsto a_i\}_{i\in I}} & = \{ x_i \mapsto \bot_{a_i} \}_{i\in I}
  \end{align*}
  \caption{Least approximation}
  \label{fig:least}
\end{minipage}
\end{figure}

An approximation value $a' : \denapproxT{A}$ is an approximation of a total
value $a : \denevalT{A}$, a relation denoted $a' \prec a$, when informally
``$a'$ has the same shape as $a$'', ignoring $\mathbf{thunk}$s, and some
subterms of $a$ may have been replaced with $\bot$ in $a'$. The set of
approximations of $a$ is defined by
$$\denapproxTT{A}{\prec a} \overset{\mathrm{def}}{=} \left\{ a' \in \denapproxT{A} \mid a' \prec a \right\}$$

The following transitivity property composes $\le$ and $\prec$ into $\prec$.

\begin{lemma}[transitivity]\label{prop:trans}
If $a' \le a''$ and $a'' \prec a$ then $a' \prec a$.
\end{lemma}

The set $\denapproxTT{A}{\prec a}$ is a semi-lattice: two approximations
$a_1 \prec a$ and $a_2 \prec a$ can be joined into a supremum $a_1 \sqcup_A a_2$
(abbreviated as $a_1 \sqcup a_2$ when the type is clear): it is the smallest
approximation of $a$ more defined than both $a_1$ and $a_2$. The join is defined
formally in \Cref{fig:join}; it is also extended to environments
$\{ x_i \mapsto a_i \}_i : \denapproxT{\Gamma}$. We remark that a meet
$a_1 \sqcap_A a_2$ could also be defined so that approximations of an element
form a lattice, but we won't need it. \lai{I'm a bit confused about this remark, but it could be my lack of background.}
\begin{lemma}[Supremum]
  For all $a_1, a_2 \prec a$,
  \begin{enumerate}
    \item \( a_1 \sqcup a_2 \prec a \)
    \item \( a_1 \le a_1 \sqcup a_2 \) and \( a_2 \le a_1 \sqcup a_2 \)
    \item for all \( a' \prec a \), \( a_1 \le a' \wedge a_2 \le a' \implies a_1 \sqcup a_2 \le a' \)
  \end{enumerate}
\end{lemma}

\begin{figure}
  \begin{align*}
    \sqcup & : \denapproxTT{A}{\prec a} \times \denapproxTT{A}{\prec a} \to \denapproxTT{A}{\prec a} \\
    \bot \sqcup_{T\,A} \bot & = \bot \\
    \bot \sqcup_{T\,A} \mathbf{thunk}\,b & = \mathbf{thunk}\,b \\
    \mathbf{thunk}\,b \sqcup_{T\,A} \bot & = \mathbf{thunk}\,b \\
    \mathbf{thunk}\,b \sqcup_{T\,A} \mathbf{thunk}\,c & = \mathbf{thunk}\,(b \sqcup_A c) \\
    \mathbf{cons}\,b\,c \sqcup_{\mathrm{list}\,A} \mathbf{cons}\,d\,e & = \mathbf{cons}\,(b \sqcup_{T\,A} d)\,(c \sqcup_{T\,(\mathrm{list}\,A)} e) \\
    \mathbf{nil} \sqcup_{\mathrm{list}\,A} \mathbf{nil} & = \mathbf{nil} \\
    \{ x \mapsto g_x \}_{(x : A)\in\Gamma} \sqcup_{\mathrm{\Gamma}} \{ x \mapsto g_x' \}_{(x : A)\in\Gamma}
      & = \{x \mapsto g_x \sqcup_{A} g_x' \}_{(x : A)\in\Gamma}
  \end{align*}
  \caption{Joining approximations (NB: cases with mismatched constructors cannot happen)}
  \label{fig:join}
\end{figure}

An element $a : \denevalT{A}$ has a \emph{least approximation} $\bot_a : \denapproxT{A}$.
This is simply $\bot$ when $A$ is of the form $T B$, but otherwise $\bot$ is not an element
of $\denapproxT{A}$, and $\bot_a$ must be the head constructor of $a$ applied to least
approximations of its fields. The least approximation is defined formally in \Cref{fig:least},
also extended to environments $\{ x_i \mapsto a_i \}_i : \denapproxT{\Gamma}$.
As its name implies, it is smaller than all other approximations of $a$.
\begin{lemma}[bottom]
  If $a' \prec a$, then $\bot_a \le a'$.
\end{lemma}

\paragraph{Backwards evaluation}
Given an input $g : \denevalT{\Gamma}$ and an approximation $a$ of the output
($a \prec \deneval{M}(g)$),
the demand semantics $\dendem{M}(g,a) : \mathbb{N} \times \denapproxTT{\Gamma}{\prec g}$
describes the cost of \emph{lazily evaluating $M$ with demand $a$},
that is, evaluating $M$ to a value $a'$ ``at least as defined as $a$,''
a relation which will be denoted by $a \le a'$.
The resulting semantics is the cost of doing that evaluation,
as well as the demand on the input, \ie{} the minimal approximation of the input $g$
that is sufficient to match the output demand $a$.
The demand semantics is defined in \Cref{fig:demand-semantics}.
Let us remark that the equation of $\dendem{}$ for $\tlet{x}{M}{N}$ contains occurrences
of both $\deneval{M}$ and $\dendem{M}$ (and similarly for $\mathrm{foldr}$).
If we view this semantics as a translation, the size of the demand function may thus
grow quadratically with respect to the size of the original function.
This effect is mitigated in practice because functions bodies are usually small.

\subsection{Properties of Demand Semantics}\label{subsec:properties}

We have seen that our semantics consists of a pair of forward and backward evaluation functions.
Let us describe a few key properties of these functions.

For the rest of this section, let $\Gamma \vdash M : A$ be a well-typed term,
$g \in \denevalT{\Gamma}$, and $a = \deneval{M}(g)$.

\begin{figure}
  \begin{tikzcd}
    g & & a \\
    g' & & a'
    \arrow["\deneval{M}",->,from=1-1,to=1-3]
    \arrow["\prec",->,dotted,from=2-1,to=1-1]
    \arrow["\prec",->,from=2-3,to=1-3]
    \arrow["{\dendem{M}(g,\cdot)}",<-,dotted,from=2-1,to=2-3]
  \end{tikzcd}
  \caption{Diagram of \Cref{prop:eval-dem} (totality). Full arrows are quantified universally.
  Dotted arrows are quantified existentially.}\label{fig:eval-dem}.
\end{figure}
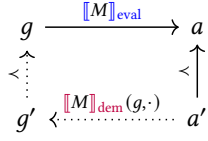

\emph{Totality} says that the demand semantics is defined
for all approximations of outputs of the pure function $\deneval{M}$.
This property is pictured as a commutative diagram in \Cref{fig:eval-dem},
where the dotted arrows are existentially quantified.
Let $a$ be an output of $\deneval{M}$
(in \Cref{fig:eval-dem}, this is represented by the arrow $g \xrightarrow{\deneval{M}} a$,
meaning that $a = \deneval{M}(g)$).
Given an approximation $a'$ of the output ($a' \le a$)
the demand semantics is defined on $a'$
(\ie{} there exists an arrow $g' \xleftarrow{\dendem{M}(g,\cdot)} a'$, meaning that
$(n, g') = \dendem{M}(g,a')$ for some $n$)
yielding an approximation of the input ($g' \le g$).
As the demand semantics is a partial function,
we write $\exists (n,g') = \dendem{M}(g,a')$ to assert
that $\dendem{M}(g,a')$ is defined.
\Cref{prop:eval-dem} expresses this property formally.

We abuse notation slightly, writing $\exists (n, g') = \dendem{M}(g,a') \wedge P$ as
shorthand for $\exists (n,g'), (n, g') = \dendem{M}(g,a') \wedge P$,
meaning that $\dendem{M}(g,a')$ is defined and its value $(n, g')$ satisfies the proposition $P$.
\begin{lemma}[Totality]\label{prop:eval-dem}
  Let $g \in \denevalT{\Gamma}$ and $a' \in \denapproxT{A}$ such that $a' \prec \deneval{M}(g)$.
\[
  \exists (n, g') = \dendem{M}(g,a') \wedge g' \prec g
\]
\end{lemma}

In \Cref{fig:demand-semantics}, $\dendem{M}$ was given a signature as a partial function.
Knowing that $\dendem{M}$ satisfies that totality property, we can indeed view it as a total function
with a type depending upon the first argument $g : \denevalT{\Gamma}$:
\[
  \dendem{\Gamma \vdash M : A} : (g : \denevalT{\Gamma}) \times \denapproxTT{A}{\prec \deneval{M}(g)} \to \mathbb{N} \times \denapproxTT{\Gamma}{\prec g}
\]

The demand semantics is monotone: the more output is demanded from $M$,
the more input it demands, and the higher cost it takes to produce the demanded output.

\begin{lemma}[monotonicity]\label{prop:mono}
  Let $g \in \denevalT{\Gamma}$ and $a_1,a_2 \in \denapproxT{A}$ such that $a_1 \le a_2 \prec \deneval{M}(g)$.
  \[
    \dendem{M}(g,a_1) \le \dendem{M}(g,a_2)
  \]
  where
  \(
    (n_1,g_1) \le (n_2,g_2) \overset{\mathrm{def}}{\iff} n_1 \le n_2 \wedge g_1 \le g_2
  \).
\end{lemma}

The demand semantics almost commutes with join ($\sqcup$). The input demand for
a union of output demands is the union of their individual input demands.
However, the cost may be less than the sum: the shared parts of the output
demands only need to be evaluated once. For example, when $a_1 = a_2$ and
$n_1 = n_2$, we have $n_1 < n_1 + n_1$ when $n_1 \neq 0$.

\begin{lemma}[$\sqcup$-homomorphism]\label{prop:homo}
  Let $g \in \denevalT{\Gamma}$ and $a_1,a_2 \prec \deneval{M}(g)$.
\[
  \dendem{M}(g,a_1 \sqcup a_2) \leqq \dendem{M}(g,a_1) \lubplus \dendem{M}(g,a_2)
\]
where $(n_1,g_1) \lubplus (n_2,g_2) \overset{\mathrm{def}}{=} (n_1 + n_2, g_1 \sqcup g_2)$ and $(n_1,g_1) \leqq (n_2,g_2) \overset{\mathrm{def}}{\iff} n_1 \le n_2 \wedge g_1 = g_2$.
\end{lemma}

\subsection{Correctness: Correspondence with Clairvoyant Semantics}\label{subsec:equiv}

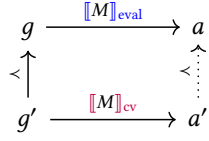
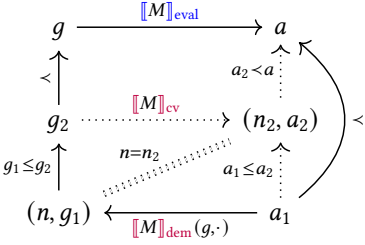
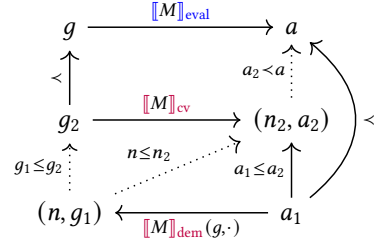
\begin{figure}
\begin{subfigure}{0.9\linewidth}
  \centering
  \begin{tikzcd}
    g & & a \\
    g' & & a'
    \arrow["\deneval{M}",->,from=1-1,to=1-3]
    \arrow["\prec",->,from=2-1,to=1-1]
    \arrow["\prec",->,dotted,from=2-3,to=1-3]
    \arrow["\dencv{M}",->,from=2-1,to=2-3]
  \end{tikzcd}
  \caption{\Cref{thm:fun-correct} (functional correctness)}
  \label{fig:fun-correct}
\end{subfigure}
\begin{subfigure}{0.45\linewidth}
  \vspace{0.7em}
  \centering
  \begin{tikzcd}
    g  & & a \\
    g_2 & & (n_2,a_2)\\
    (n,g_1) & & a_1
    \arrow["\deneval{M}",->,from=1-1,to=1-3]
    \arrow["\prec",from=2-1,to=1-1]
    \arrow["a_2\prec a",dotted,from=2-3,to=1-3]
    \arrow["{\dencv{M}}",->,dotted,from=2-1,to=2-3]
    \arrow["{g_1 \le g_2}",->,from=3-1,to=2-1]
    \arrow["a_1 \le a_2",dotted,from=3-3,to=2-3]
	  \arrow["\prec"',curve={height=30pt}, from=3-3, to=1-3]
    \arrow["{\dendem{M}(g,\cdot)}",->,from=3-3,to=3-1]
    \arrow["n=n_2",Rightarrow,no head,dotted,from=3-1,to=2-3]
  \end{tikzcd}
  \caption{\Cref{thm:cost-existence} (cost existence)}
  \label{fig:cost-existence}
\end{subfigure}
\begin{subfigure}{0.45\linewidth}
  \centering
  \begin{tikzcd}
    g  & & a \\
    g_2 & & (n_2,a_2) \\
    (n,g_1) & & a_1
    \arrow["\deneval{M}",->,from=1-1,to=1-3]
    \arrow["\prec",from=2-1,to=1-1]
    \arrow["a_2\prec a",dotted,from=2-3,to=1-3]
	  \arrow["\prec"',curve={height=30pt}, from=3-3, to=1-3]
    \arrow["{\dencv{M}}",->,from=2-1,to=2-3]
    \arrow["n \le n_2",dotted,from=3-1,to=2-3]
    \arrow["{g_1 \le g_2}",->,dotted,from=3-1,to=2-1]
    \arrow["a_1 \le a_2",from=3-3,to=2-3]
    \arrow["{\dendem{M}(g,\cdot)}",->,from=3-3,to=3-1]
  \end{tikzcd}
  \caption{\Cref{thm:minimality} (cost minimality)}
  \label{fig:minimality}
\end{subfigure}
\caption{Diagrams of correctness theorems between demand semantics and clairvoyant semantics}
\end{figure}

Our framework relies on interpreting lazy programs as demand functions.
\citet{bjerner1989} introduced demand semantics in an untyped setting, but they
did not prove a correspondence with any other semantics. Here, we formally
relate that semantics to \emph{clairvoyant
  semantics}~\citep{forking-paths,clairvoyant}, a forward and nondeterministic
semantics that was previously related to the (operational) \emph{natural
  semantics} of laziness~\citep{launchbury1993} by \citeauthor{clairvoyant}.
\ifextended Compared with the demand semantics, the clairvoyant semantics works
forward, but uses nondeterminism to generate branches for all potential future
demands. \fi

The monadic clairvoyant semantics of~\citet{forking-paths} is
denoted $\dencv{\Gamma \vdash M : A} : \denapproxT{\Gamma} \to \mathcal{P}(\mathbb{N}\times \denapproxT{A}) $.
Note a minor difference from our calculus to the one in~\citet{forking-paths}:
the present calculus is essentially a call-by-value calculus with an explicit thunk type,
whereas~\citet{forking-paths} defines a calculus with laziness by default,
introducing thunks in the denotation of types.
Intuitively, the clairvoyant semantics of \citet{forking-paths} can be decomposed into
an elaboration to the present calculus, followed by its clairvoyant semantics
$\dencv{M}$.
The calculus presented here corresponds more
closely to the core combinators of the clairvoyance monad in~\citet{forking-paths}.

Functional correctness says that the clairvoyant semantics $\dencv{M}$
approximates the pure function $\deneval{M}$.
This theorem is pictured as a commutative diagram in \Cref{fig:fun-correct}.
Let $g$ be an environment, let $a = \deneval{M}(g)$
(diagrammatically: $g \xrightarrow{\deneval{M}} a$),
let $g' \prec g$, and let $a'$ be a nondeterministic output of $\dencv{M}(g')$
(diagrammatically: $g' \xrightarrow{\dencv{M}} a'$; formally: $(n, a') \in \dencv{M}(g')$
for some $n$). Then $a'$ approximates $a$.

\begin{theorem}[Functional correctness]\label{thm:fun-correct}
  Let $g \in \denevalT{\Gamma}$, and $g' \prec g$.
\[
  \forall (n, a') \in \dencv{M}(g'),\; a' \prec \deneval{M}(g)
\]
\end{theorem}

The demand semantics $\dendem{M}(g, a_1)$ finds a minimal pair $(n, g')$ such that
to produce a result at least as defined as the output demand $a_1$,
the clairvoyant semantics must be applied to an input at least as defined as $g'$
and the associated cost will be at least $n$.
The minimality of $(n,g')$ can formalized as the conjunction of an existence property
(the minimal cost is achievable) and a universality property (all other candidate
executions have a higher cost). Those theorems are illustrated diagrammatically in
\Cref{fig:cost-existence} and \Cref{fig:minimality}.

\begin{theorem}[Cost existence]
  \label{thm:cost-existence}
  Let $g \in \denevalT{\Gamma}$, $a_1 \prec \deneval{M}(g)$,
  let $(n, g_1) = \dendem{M}(g, a_1)$, and let $g_2$ such that $g_1 \le g_2 \prec g$.
  \[
    g_1 \le g_2 \implies \exists (n_2, a_2) \in \dencv{M}(g_2),\; n = n_2 \wedge a_1 \le a_2
  \]
\end{theorem}

\begin{theorem}[Cost minimality]
  \label{thm:minimality}
  Let $g \in \denevalT{\Gamma}$, $a_1 \prec \deneval{M}(g)$,
  let $(n, g_1) = \dendem{M}(g, a_1)$, and let $g_2 \prec g$.
  \[
    \forall (n_2, a_2) \in \dencv{M}(g_2),\; a_1 \le a_2 \implies n \le n_2 \wedge g_1 \le g_2
  \]
\end{theorem}

\subsection{Deriving the Definition of $\dendem{M}$}

We can use the properties above to derive the definition of $\dendem{M}$
by inequational reasoning.
For example, to find the value of $\dendem{\mathrm{force}\,M}$ as a function of
$\dendem{M}$, totality requires $\dendem{\mathrm{force}\,M}(g,a') \prec g$,
given $a' \prec_{A} \deneval{\mathrm{force}\,M}(g) = \deneval{M}(g)$,
and given the totality property for $M$, $\dendem{M}(g,a'') \le g$ for all $a'' \prec_{T\,A} a$.
With $a'' = \mathbf{thunk}\,a'$, we have $\dendem{M}(g,\mathbf{thunk}\,a') \le g$.
This suggests the definition $\dendem{\mathrm{force}\,M}(g,a') = \dendem{M}(g,\mathbf{thunk}\,a')$.
In a similar way, we can derive the tricky-looking definition of
$\dendem{\mathrm{foldr}\,M_1\,M_2\,N}$
by inequational reasoning.

Note that the minimality property forbids the trivial definition
$\dendem{M}(g,a)=(0,\bot_g)$.

\subsection{Limitations of This Work}\label{subsec:limitations}

All of our examples of demand functions were translated manually.
Although the demand semantics~(\Cref{sec:demand-semantics})
could be used to systematically translate from a pure function to a demand function,
further efforts are necessary to simplify the generated code into a readable result
conducive to mechanical reasoning. An automatic translation would significantly
improve the usability of our framework.

Nevertheless, as manual translations can be error-prone, we have developed a
method to ensure our translation is correct by cross-validating it with
the clairvoyant semantics~\citep{clairvoyant, forking-paths}. We will explain this
framework in more detail in \cref{sec:case-study-1} and
\cref{sec:case-study-2}.

The bidirectional demand semantics presented here does not support general
recursive and higher-order functions, which limits the expressiveness of the
demand semantics. However, our case studies show that even with those
restrictions, the demand semantics is still useful for mechanically reasoning
about the time cost of many lazy functions as well as amortized and persistent
data structures.

The lack of general recursion is not restrictive in the context of time
complexity analysis. We can simulate general recursion using a fuel parameter
which decreases with every recursive call. This does not change the asymptotic
complexity of programs, and time complexity bounds will determine how much fuel
is enough.

Many classical data structures and algorithms in complexity analysis
only involve first-order functions such as \cdc{push} and \cdc{pop} for queues.
That is what our present work focuses on.
It would be useful to support higher-order functions to reason about operations
such as maps, traversals, filters, and folds.
The bidirectional demand semantics originally presented by~\cite{bjerner1989}
does not feature higher-order functions either. Extending those semantics with
higher-order functions would complicate the results of this section. Currently,
we use the same denotation of types ($\denapproxT{A}$) for both the demand
semantics and the clairvoyant semantics, allowing us to easily compare values
between the two semantics. In the clairvoyant semantics, the interpretation of function
types $A \to B$ as monadic functions
$\denapproxT{A} \to \mathcal{P}(\mathbb{N}\times \denapproxT{B})$
is specific to clairvoyant semantics.
It does not seem suitable to represent the demand on a function.
Intuitively, functions propagate demand in two ways. When a function is applied,
the demand on its result is mapped to the demand on its arguments and the function itself.
When a function is defined, the demand on the function is mapped to a demand on
the context of the lambda abstraction. It would be interesting to find a representation
of demand for functions that meets those needs.

Demand semantics let us reason about total time cost, but not real-time
(\ie{} non-amortized) cost, or space cost.
This is because demand functions only calculate \emph{if} a thunk
is evaluated, not \emph{when} a thunk is evaluated.

\section{Case Studies: Sorting Algorithms}\label{sec:case-study-1}

In this section, we analyze insertion sort and selection sort using our
bidirectional demand semantics to demonstrate how our model can be used in
practice. These algorithms are known to exhibit $O(n^{2})$ time complexity under
eager evaluation. However, we can achieve $O(k \cdot n)$ complexity under lazy
evaluation if we only need the smallest $k$ elements of a list. More formally,
given the following functions:
\begin{lstlisting}[language=Coq]
Definition p1 (k : nat) (xs : list nat) := take k (insertion_sort xs).
Definition p2 (k : nat) (xs : list nat) := take k (selection_sort xs (length xs)).
\end{lstlisting}
We will prove that the computation cost of both \cdc{p1 k xs} and \cdc{p2 k xs}
are bounded by $O(k \cdot n)$ where $n=|\texttt{xs}|$. All the lemmas and
theorems we show in this section have been formally proven in the Rocq Prover.\@
This code can be found in our artifact~\citep{demand-artifact}.

\subsection{The \texorpdfstring{\cdc{take}}{take} Function}\label{subsec:take}

We show the definitions of \cdc{take} and \cdc{takeD} in \cref{fig:take}.
Surprisingly, \cdc{take} is not expressible in the calculus defined in the
previous section: indeed, to define \cdc{take} using \cdc{foldr}, we also need
first-class functions. We were not able to remove this unfortunate limitation without
significantly increasing the complexity of our calculus. In practice, we are
still able to manually define a demand function for \cdc{take} and many similar functions.
As a safeguard, we cross-validate this demand function \cdc{takeD} with the clairvoyant
semantics, which is simply a monadic translation \cdc{takeA}.
We prove a correspondence between \cdc{takeD} and \cdc{takeA}
in the form of the theorems of \Cref{subsec:equiv}.

To define \cdc{takeD}, we apply
$\dendem{\cdot}(g, d)$~(\cref{fig:demand-semantics}) to the definition of
\cdc{take} to obtain the following:
\begin{align*}
  \dendem{\mathrm{foldr}\,(\lambda (n, y)\, zs.\, \mathbf{cons}\,y\,zs)\,\mathbf{nil}\,t}(g, d) & = (c, g') \lubplus \dendem{t}(g, n') \\
  \text{where}\,(c, g', n') & = \mathbf{foldr}_{\demcolor\mathrm{dem}}(g,\mathbf{cons}\,y\,zs,\mathbf{nil},\mathbf{thunk}\,(\deneval{t}(g)),\mathbf{thunk}\,d)
\end{align*}
Because $t$ is an argument of function \cdc{takeD} so we use the rule for
evaluating variables to get $\dendem{t}(g, n') = (0, \{t \mapsto n'\})$.
However, before we move on, we would like to treat specially the numbers
contained in argument $t$. Even though we use the inductive \cdc{nat} data type
in Gallina to represent numbers for simplicity, numbers are primitive data types
in most programming languages. As such, we are not interested in an
approximation of a \cdc{nat}, so we make a simplification in \cdc{takeD} such
that we only return the demand of the argument \cdc{xs}. For this reason, we
will ignore the numbers in $t$ in the rest of the translation process.

\begin{figure}[t]
\begin{lstlisting}[language=Coq,numbers=left,xleftmargin=4.0ex]
Fixpoint take {A} (k : nat) (xs : list A) : list A :=
  match k, xs with
  | O, _ => nil
  | S _, nil => nil
  | S k', x :: xs' =>
    let zs := take k' xs' in
    x :: zs
  end.

Fixpoint takeD {A} (k : nat) (xs : list A)
                   (outD : listA A) : Tick (T (listA A)) :=
  tick >> match k, xs, outD with
  | O, _, _ => ret Undefined
  | _, nil, _ => ret (Thunk NilA)
  | S k', x :: xs', ConsA zD zsD =>
    let+ ysD := thunkD (takeD k' xs') zsD in
    ret (Thunk (ConsA (Thunk x) ysD))
  | _, _, _ => bottom (* absurdity case *)
  end.
\end{lstlisting}
\caption{The Gallina implementation of \cdc{take} and
  \cdc{takeD}.}\label{fig:take}
\end{figure}

The $\mathbf{foldr}_{\demcolor\mathrm{dem}}$ function is a recursive function
that supports pattern matching~(\cref{fig:foldr-dem}). It lives in the universe
of denotations that we would like to replace with Gallina definitions---in fact,
this will be our definition of \cdc{takeD}. In the case that $t = \mathbf{nil}$, we
return $\mathbf{thunk}\,\mathbf{nil}$~(line~14, \cref{fig:take}). In the
case that $t = \mathbf{cons}\,(n, y)\,ts$, there are two steps. First, we
need to run the following:
\begin{align*}
  (c_1,\{g_1,y \mapsto a_1',zs\mapsto b_2'\}) = &\dendem{M_1}(\{g, y\mapsto a_1,zs\mapsto \mathbf{foldr}_{\evalcolor\mathrm{eval}}(g,M_1,M_2, a_2)\}, d) \\
                                                   = &\dendem{\mathbf{cons}\,y\,zs}(\{g,y\mapsto y,\\
                                                                                        &\qquad zs\mapsto \mathbf{foldr}_{\evalcolor\mathrm{eval}}(g,\mathbf{cons}\,y\,zs,\mathbf{thunk}\,\mathbf{nil}, zs)\}, d)
\end{align*}
However, this expression only makes sense when $d = \mathbf{cons}\,zD\,zsD$
according to \cref{fig:demand-semantics}, so we perform a pattern match on $d$ as
well (line~15). We will handle $d = \mathbf{nil}$ as one of the absurdity cases
(line~18). If we continue running the demand semantics, we will see
that this part evaluates to 
$(c_1,\{g_1,y \mapsto a_1',zs\mapsto b_2'\}) = (0, \{y\mapsto zD, zs\mapsto zsD\})$. After that, we need to run:
\begin{align*}
  (c_2, g_2, a_2') = &\mathbf{foldr}_{\demcolor\mathrm{dem}}(g,M_1,M_2,a_2,b_2') \\
  = &\mathbf{foldr}_{\demcolor\mathrm{dem}}(g,\mathbf{cons}\,y\,zs,\mathbf{nil},zs,zsD)
\end{align*}
This is equivalent to a recursive call applied to $zsD$ (line~16). According to
the definition of $\mathbf{foldr}_{\demcolor\mathrm{dem}}$, we obtain
$(c_1 + c_2, g_1 \sqcup g_2, \mathbf{thunk}\,(\mathbf{cons}\,a_1'\,a_2'))$.
Finally, combining this result with the denotational semantics, we obtain the
cost $c_1 + c_2$ and the input demand
$\mathbf{thunk}\,(\mathbf{cons}\,a_1'\,a_2')$. As explained previously, we use
the \cdc{Tick} data type to represent the tuple of computation cost and input
demand. It is defined as a writer monad such that costs are added together in a
monadic bind~(the \cdc{let+} notation on line~16). Thanks to the use of monads,
we only need to \cdc{ret} the input demand~(line~17).

After these steps, we need to add back the \cdc{nat} argument \cdc{k} to the
definition. In the case that \cdc{k = O}, the original \cdc{take} function
returns \cdc{nil}. By the rule for \cdc{nil} in \cref{fig:demand-semantics}, we
know that we should return \cdc{Undefined}~($\bot$) in this case~(line~13). We
also handle the case in which the output demand is $\bot$ and the case in which
we apply $\dendem{\mathbf{cons}\,y\,zs}$ to a $d$ that is not a $\mathbf{cons}$.
In these cases, we return \cdc{bottom}, which represents a cost of 0 and the
minimal input demand, \ie~\cdc{Undefined} in this case~(line~18).

Finally, we manually add a \cdc{tick} in the beginning of the function~(line~12)
so that we can count the number of function calls.


\paragraph{Cost} Now that we have defined \cdc{takeD}, we can use it to reason
about the cost of \cdc{take}. We can state and prove the following cost
theorems:
\begin{lstlisting}[language=Coq]
Theorem takeD_cost : forall {A : Type} (n : nat) (xs : list A) outD,
    Tick.cost (takeD n xs outD) <= 1 + n.

Theorem takeD_cost' : forall {A : Type} (n : nat) (xs : list A) outD,
    Tick.cost (takeD n xs outD) <= sizeX' 1 outD.
\end{lstlisting}
Both theorems define aspects of the cost of a lazy \cdc{take}. The function
\cdc{Tick.cost} projects the computation cost from the \cdc{Tick} monad. The
first theorem states that the cost is bounded by its parameter \cdc{n} + 1. The
second theorem states that the cost is also bounded by the size of the output
demand. The function \cdc{sizeX'} is defined such that \cdc{sizeX' 1 outD} is
equivalent to $\text{max}(1, |\hbox{\cdc{outD}}|)$.

\paragraph{Functional correctness} Since our translation is manual, we
additionally prove the following functional correctness theorem for \cdc{takeD}:
\begin{lstlisting}[language=Coq]
Lemma takeD_approx (n : nat) (xs : list nat) outD :
  outD `is_approx` take n xs ->
  Tick.val (takeD n xs outD) `is_approx` xs.
\end{lstlisting}
The function \cdc{Tick.val} projects the value from the \cdc{Tick} monad. We
prove a similar theorem for every demand function that we translate so that we
can be more confident about our translation.

\subsection{Insertion Sort}\label{subsec:insertion-sort}

We have shown the pure implementation of the insertion sort and its
corresponding demand functions in \cref{fig:take-insertion-sort} and
\cref{fig:insertD}, respectively. The process of translating \cdc{insert} is
similar to translating \cdc{take} except for the need to translate an
\cdc{if}-expression. Translating \cdc{insertion_sort}, however, is more
challenging. Doing so involves translating \cdc{let}-expressions with function
calls. According to the \cdc{let}-rule of the demand
semantics~(\cref{fig:demand-semantics}), we first need to run \cdc{insert} in
the forward direction, then use its result to run \cdc{insertion_sort} in the
backward direction~(\ie~\cdc{insertion_sortD}), and finally use the input demand
from \cdc{insertion_sortD} to run \cdc{insert} in the backward
direction~(\ie~\cdc{insertD}). This process corresponds to our illustration in
\cref{fig:insertion_sortD_illustration}.

\begin{figure}
\begin{lstlisting}[language=Coq]
(* Computation cost. *)
Theorem insertD_cost x (xs : list nat)  (outD : listA nat) :
  Tick.cost (insertD x xs outD) <= leb_count x xs + 1.
Theorem insertD_cost' x (xs : list nat) (outD : listA nat) :
  Tick.cost (insertD x xs outD) <= sizeX' 1 outD.
Theorem insertD_cost'' x (xs : list nat) (outD : listA nat) :
  Tick.cost (insertD x xs outD) <= length xs + 1.
Theorem insertion_sortD_cost (xs : list nat) (outD : listA nat) :
  Tick.cost (insertion_sortD xs outD) <= (sizeX' 1 outD + 1) * (length xs + 1).
Theorem selectD_cost (x : nat) (xs : list nat) (yD : nat) (ysD : listA nat) :
  Tick.cost (selectD x xs (pairA yD ysD)) <= length xs + 1.
Theorem selection_sortD_cost (xs : list nat) (n : nat) (outD : listA nat) :
  n >= length xs ->
  Tick.cost (selection_sortD xs n outD) <= (sizeX' 1 outD) * (length xs + 1).

(* Functional correctness. *)
Theorem insertD_approx (x : nat) (xs : list nat) (outD : listA nat) :
  outD `is_approx` insert x xs ->
  Tick.val (insertD x xs outD) `is_approx` xs.
Theorem insertion_sortD_approx (xs : list nat) (outD : listA nat) :
  outD `is_approx` insertion_sort xs ->
  Tick.val (insertion_sortD xs outD) `is_approx` xs.
Theorem selectD_approx (x : nat) (xs : list nat) (outD : prodA nat (listA nat)):
  outD `is_approx` select x xs ->
  Tick.val (selectD x xs outD) `is_approx` xs.
Theorem selection_sortD_approx (xs : list nat) (n : nat) (outD : listA nat) :
  outD `is_approx` selection_sort xs n ->
  Tick.val (selection_sortD xs n outD) `is_approx` xs.

(* Composition. *)
Theorem take_insertion_sortD_cost (n : nat) (xs : list nat) (outD : listA nat) :
  Tick.cost (take_insertion_sortD n xs outD) <= (n + 1) * (length xs + 2) + 1.
Theorem take_selection_sortD_cost (n : nat) (xs : list nat) (outD : listA nat) :
  Tick.cost (take_selection_sortD n xs outD) <= n * (length xs + 2) + 1.
\end{lstlisting}
\caption{Main theorems we have proven for both insertion
  sort and selection sort.}\label{fig:insertion-theorems}\label{fig:selection-theorems}
\end{figure}

We show the main theorems we have proven for insertion sort in
\cref{fig:insertion-theorems}. Proofs for these theorems are relatively
straightforward. Theorems regarding \cdc{insertD} and \cdc{insertion_sortD} can
all be proven by an induction over the list arguments \cdc{xs}. All the
inequalities involved in these theorems can be solved by Rocq Prover's built-in
tactics \cdc{lia} and \cdc{nia}~\citep{micromega}. We also defined our custom
own tactics to help solve \cdc{is_approx} relations.

In addition, we combine the theorems \cdc{takeD_cost} and
\cdc{insertion_sortD_cost} to prove the theorem \cdc{take_insertion_sortD_cost}
~(\cref{fig:insertion-theorems}). The function \cdc{take_insertion_sortD} is the
demand function of \cdc{take_insertion_sort}, which composes \cdc{take} and
\cdc{insertion_sort}. By proving this theorem, we formally prove that the cost
of \cdc{take_insertion_sortD k xs outD} is bounded by $O(k \cdot n)$ where
$n = |\hbox{\cdc{xs}}|$.

Theoretically, we can show a tighter bound in \cdc{take_insertion_sortD_cost},
as the list argument for \cdc{insertion_sort} decreases after each recursive
call. However, the bound we show here is sufficient to show that the function
has $O(k \cdot n)$ time complexity. Proving this tighter bound would not change
the asymptotic cost.

\subsection{Selection Sort}\label{subsec:selection-sort}

The computation cost of \cdc{selection_sort} is bounded by the same time
complexity as \cdc{insertion_sort}. To show that, we take the implementation of
\cdc{select} and \cdc{selection_sort} from \emph{Verified Functional
  Algorithms}~\citep{vfa}, then manually translate them into demand functions.
For the sake of space, we only show the types of all relevant definitions in
\cref{fig:selection_sort}. In addition to the list argument \cdc{xs}, the
\cdc{selection_sort} function takes an additional argument \cdc{n : nat}, which
is our ``fuel'' for running selection sort. We use this fuel to convince the
Gallina termination checker that \cdc{selection_sort} will
terminate.\footnote{It is possible to manually prove that \cdc{selection_sort}
  terminates without using this ``fuel'' construct. We demonstrate the fuel
  version here for simplicity, as it requires fewer proofs.} In practice, we
always want to use a fuel size that is at least as large as the length of
\cdc{xs}. The complete definitions and proofs for selection sort are included in
our artifact.

\begin{figure}[t]
\begin{lstlisting}[language=Coq]
Definition select (x: nat) (l: list nat) : nat * list nat.
Definition selection_sort (l : list nat) (n : nat) : list nat.
Definition selectD (x : nat) (l : list nat)
  (outD : prodA nat (listA nat)) : Tick (T (listA nat)).
Definition selection_sortD (l : list nat) (n : nat)
  (outD : listA nat) : Tick (T (listA nat)).
\end{lstlisting}
\caption{Type signatures of all definitions related to the selection
  sort.}\label{fig:selection_sort}
\end{figure}

Compared to \cdc{insertion_sort}, \cdc{selection_sort} is more challenging
because it returns a product type. Accordingly, we need to use \cdc{prodA A B =
  T A * T B}, an approximation of the product type, as the type of \cdc{outD} in
\cdc{selectD}. This typing is crucial as, unlike \cdc{insert}, the cost of
\cdc{select} is not bounded by the demand on the output list. In fact, even when
the demand on the output list is \cdc{Undefined}, as long as the demand on the
output number (the smallest number in the list) is not \cdc{Undefined},
\cdc{select} still must traverse the entire input list to select such a number.

We show the main theorems we have proven for selection sort in
\cref{fig:selection-theorems}. Compared to \cdc{insertD} in insertion sort,
\cdc{selectD} is only bounded by \cdc{length xs + 1}, not \cdc{sizeX' 1 outD},
because the function always traverses the entire input argument \cdc{xs}.
Nevertheless, we show a similar bound for \cdc{selection_sortD} as long as the
fuel we use is greater than or equal to \cdc{length xs}. We also additionally
prove the functional correctness theorems of \cdc{selectD} and
\cdc{selection_sortD} to validate our manual translation. In the end, we can
compose \cdc{takeD_cost} with \cdc{selection_sortD_cost} to prove the bounds
stated in \cdc{take_selection_sortD_cost}.\footnote{The
  \cdc{take_selection_sort} function runs \cdc{selection_sort} with a fuel equal
  to \cdc{length xs}.}

\section{Amortized and Persistent Data Structures}\label{sec:example}\label{sec:physicist}\label{sec:case-study-2}

In this work, we apply our method to mechanically reason about two \emph{lazy},
\emph{amortized}, and \emph{persistent} data structures,
\Citeauthor{purely-functional}'s banker's queue and implicit
queue~\citep{purely-functional}. Both data structures implement
first-in-first-out~(FIFO) queues with amortized constant time operations. The
banker's queue achieves amortization and persistence by maintaining a balancing
invariant on two lists. The implicit queue achieves both properties using a
technique called ``implicit recursive slowdown''~\citep{purely-functional,
  recursive-slowdown}.

\ifextended
Our amortized and persistent analysis is based on a novel \emph{\method} that is
designed for our demand semantics~(\cref{subsec:method}). We manually derive the
demand functions and prove that both the banker's queue and the implicit queue
are amortized and persistent using the Rocq
Prover~(\cref{subsec:banker}--\ref{subsec:implicit}). The demand functions for
these data structures (especially the implicit queue) are more complex than what
was shown in \cref{sec:case-study-1}, so in our mechanized proofs, we do
\emph{not} trust our demand functions. Instead, we validate that our demand
functions are correct by corresponding them with an alternative and more
mechanical semantics called the clairvoyant semantics~\citep{forking-paths}. All
the definitions and proofs can be found in our artifact~\citep{demand-artifact}.
\fi

\subsection{The \MethodTitle}\label{subsec:method}

The banker's method and the physicist's method are the two classical methods for
analyzing amortized computation costs~\citep{amortized}. However, these methods
only work for ``forward'' and strict semantics, where we \emph{first} accumulate
credits~(in the banker's method) or potential~(in the physicist's method) before
making an ``expensive'' operation to spend the accumulation. Our demand
semantics works differently: we compute a minimal input demand from an
output demand, working ``backwards''. Therefore, we propose a new method, called
the \emph{\method}, to analyze amortized computation cost based on this
semantics.

The key idea of the \method{} is to consider the demand semantics as an
evaluation on approximations that happen ``backward''. Under this view, future
operations are accumulating \potential{} that is going to be used by expensive
operations that happen earlier. Therefore, our solution is to assign
\emph{\potential{}} to demands. Taking the banker's queue as an example, we
assign a \potential{} $\Phi$ to each demand of a queue
$q^{D}=\{nf, f^{D}, nb, b^{D}\}$ as:
\begin{equation}\label{eqn:potential}
\Phi(\{nf, f^{D}, nb, b^{D}\}) = \text{max}(2 \times (|f^{D}| - nb), 0)
\end{equation}
We use $f^{D}$ to represent the demand of the front list and $nb$ to represent
the length of the back list. We overload the $|\cdot|$ operator to represent the
length of a demand of a list, so $|f^{D}|$ represents the length of $f^{D}$. We
use $nb$ instead of $|b^{D}|$ in the potential function because moving a
\cdc{back} list to the \cdc{front} list requires reversing the entire \cdc{back}
list, regardless of the length of $b^{D}$. Note that $|f^{D}| \leq nf$ and
$|b^{D}| \leq nb$.

For each operation of the a queue~(\ie~\cdc{push} and \cdc{pop}), we show that
the following inequality holds:
\begin{equation}\label{eqn:costspec}
  \mathit{cost} \leq \Phi(q^{D}_{out})-\Phi(q^{D}_{in}) + \mathit{const}
\end{equation}
$\Phi(q^{D}_{in})$ and $\Phi(q^{D}_{out})$ represent the \potential{} for the
input queue and the output queue of the operation, respectively. The
$\mathit{cost}$ is bounded by the difference between the input demand's
potential and the output demand's potential plus a constant number
($\mathit{const}$).

\begin{figure}[t]
\begin{lstlisting}[language=Coq]
let q0 = empty in        
let q1 = push q0 a in    
let q2 = push q1 b in    (* D q2@2 = (a:nil,b:bot)                        *)
let q3 = push q2 c in    (* D q2@3 = (a:nil,b:bot)                        *)
let q4 = push q3 d in    (* D q2@4 = (bot, bot)    D q4@4 = (a:b:bot,bot) *)
(_, q5) <- pop q4 ;;     (* D q2@5 = (bot, bot)    D q4@5 = (a:b:bot,bot) *)
(_, q6) <- pop q5 ;;     (* D q2@6 = (bot, bot)    D q4@6 = (a:bot,bot)   *)
(_, q7) <- pop q4 ;;     (* D q2@7 = (bot, bot)    D q4@7 = (bot,bot)     *)
(* ... *)
\end{lstlisting}
\caption{A program that uses the banker's queue, with demands of \cdc{q2} and \cdc{q4} labeled at each step.}\label{fig:prog-queue}
\end{figure}

To show that a queue is amortized and persistent over arbitrary program traces,
we consider the demands for all versions of the queue generated in program
points. We use $q^{D}_{i@j}$ to denote the demand for $i$-th queue at the $j$-th
operation (both indices start from 0). We constrain $j$ to be at least as large
as $i$ for any $q^{D}_{{i@j}}$. Note that we do not put any constraints on the
program trace---one version of a queue can be reused for any number of times in
a program trace.

Taking the program shown in \cref{fig:prog-queue} as an example. We start from
$q^{D}_{i@j} = (\bot, \bot)$ for any $j \ge 8$, because there is no more demand
after the program has finished. Using the initial list of $q^{D}_{i@8}$ for all
$i$s as the output demand, we can compute the minimal input demands for every
queue at each step using our demand semantics~(\cref{sec:demand-semantics}).
\ifextended We obtain the following demands:
\begin{align*}
  & q^{D}_{7@j} = (\bot, \bot)  &\qquad
  & q^{D}_{6@j} = (\bot, \bot) \\
  & q^{D}_{5@j} = \begin{cases}
              (b : \bot, \bot) & \text{if } j \le 6, \\
              (\bot, \bot) & \text{otherwise}
            \end{cases} &\qquad
  & q^{D}_{4@j} = \begin{cases}
              (a : b : \bot, \bot) & \text{if } j \le 5, \\
              (a : \bot, \bot) & \text{if } 5 < j \le 7, \\
              (\bot, \bot) & \text{otherwise}
            \end{cases} \\
  & q^{D}_{3@j} = \begin{cases}
              (a : b : \bot, \bot) & \text{if } j \le 4, \\
              (\bot, \bot) & \text{otherwise}
            \end{cases} &\qquad
  & q^{D}_{2@j} = \begin{cases}
              (a : \mathbf{nil}, b : \bot) & \text{if } j \le 3, \\
              (\bot, \bot) & \text{otherwise}
            \end{cases} \\
  & q^{D}_{1@j} = \begin{cases}
              (a : \mathbf{nil}, \bot) & \text{if } j \le 2 \\
              (\bot, \bot) & \text{otherwise}
            \end{cases} &\qquad
  & q^{D}_{0@j} = (\mathbf{nil}, \mathbf{nil})
\end{align*}
\else The demands of \cdc{q2} and \cdc{q4} at each operation are shown as
comments in \cref{fig:prog-queue}. \fi

This notion allows us to lift inequality over a single operation,
\ie~Inequality~(\ref{eqn:costspec}), to the following inequality over parts of a
program trace:
\begin{equation}\label{eqn:interval-costspec}
cost_{[i, j]} \leq \sum_{k=0}^{j}{\Phi(q^{D}_{k@j})}-\sum_{k=0}^{i}\Phi(q^{D}_{k@i}) + (j - i) \cdot const
\end{equation}
We use $cost_{[i,j]}$ to represent the computation cost incurred from the $i$-th
operation to the $j$-th operation. When $j = i + 1$, $cost_{[i, j]}$ is the cost
for a single operation. In addition, we compute the difference between \emph{the
  sum} of all the \potential{} of all queue demands at step $j$ and that at step
$i$.

If we suppose that our program has $t$ operations in total, then the cost of the
entire program is $cost_{[0, t]}$. At the beginning of a program, the empty
queue is the only possible queue, and its demand is $q^{D}_{0@0} = \bot$ and
$\Phi(q^{D}_{0@0}) = 0$. At the end of a program, there cannot be any more
demands, so $q^{D}_{i@t} = \bot$ for all $0 \le i < t$ and
$\sum_{i=0}^{t-1}{\Phi(q^{D}_{i@t})} = 0$. Therefore, if we can show that
Inequality~(\ref{eqn:interval-costspec}) is true for all $i$ and $j$ in one
program trace, we can conclude that the cost of the entire program is bounded by
constant cost multiplied by the number of queue operations.

We show our Rocq Prover formalization of
Inequality~(\ref{eqn:interval-costspec}) in lines 1--5 of
\cref{fig:coq-physicist}. The definition \cdc{Physicist'sArgumentD} is defined
generally so that we can use it on both the banker's queue and the implicit
queue. The definition states that for any list of values \cdc{vs}~(line~2), if
we have a ``stack'' of approximations \cdc{output} of running \cdc{vs} using the
pure function of operation \cdc{o}, where \cdc{o} can be either a \cdc{push} or
a \cdc{pop}~(line~3), and if we can run the demand function of operation \cdc{o}
on \cdc{vs} and \cdc{output} to obtain a computation cost \cdc{cost} and an
input demand \cdc{input}~(line~4), then the sum of all \cdc{input}'s potential
plus \cdc{cost} is smaller than or equal to a budget assigned to operation
\cdc{o} over value \cdc{vs} plus the sum of all \cdc{output}'s
potential~(line~5).

\begin{figure}[t]
\begin{lstlisting}[language=Coq,numbers=left,xleftmargin=4.0ex]
Definition Physicist'sArgumentD : Prop :=
  forall (o : op) (vs : list value), well_formed vs ->
  forall output : stackA, output `is_approx` eval o vs ->
  forall input cost, Tick.MkTick cost input = demand o vs output ->
  sumof potential input + cost <= budget o vs + sumof potential output.
  
Definition AmortizedCostSpec : Prop :=
  forall os : trace, (cost_of (exec_trace os) <= budget_trace os).
\end{lstlisting}
\caption{Definitions of the \method{}~(\cdc{Physicist'sArgumentD}) as well as
  amortization and persistence~(\cdc{AmortizedCostSpec}) formalized in the Rocq
  Prover.}\label{fig:coq-physicist}
\end{figure}

The final theorem that states a queue is both amortized and persistent is shown
in lines 7--8 in \cref{fig:coq-physicist}. We show that for any queue \cdc{q},
if we can show that a queue and all its operations satisfy
\cdc{Physicist'sArgumentD}, it implies that the queue also satisfies
\cdc{AmortizedCostSpec}. In this way, we only need to show that the banker's
queue and the implicit queue both satisfy \cdc{Physicist'sArgumentD}.

\subsection{Banker's Queue}\label{subsec:banker}

We show a Gallina implementation of the banker's queue in
\cref{fig:banker-queue}. We first declare a record \cdc{Queue} whose internal
representation is two lists and two numbers: a \cdc{front} list and a \cdc{back}
list, with two numbers \cdc{nfront} and \cdc{nback} that keep track of the
lengths of these two lists, respectively~(lines 1--6). When \cdc{push}ing an
element to the queue, we add it to the head of the \cdc{back} list~(lines
13--14). When \cdc{pop}ing an element from the queue, we remove it from the head
of the \cdc{front} list~(lines 16--20). Both functions use a smart constructor
\cdc{mkQueue} to maintain the ``balance'' of the queue, by reversing and moving
the \cdc{back} list to the end of the \cdc{front} list when \cdc{front} is
shorter than \cdc{back}~(lines 8--11).

\begin{figure}[t]
\begin{lstlisting}[language=Coq,numbers=left,xleftmargin=4.0ex]
Record Queue (A : Type) : Type := MkQueue
  { nfront : nat
  ; front : list A
  ; nback : nat
  ; back : list A
  }.

Definition mkQueue {A : Type} 
  (nf : nat) (f : list A) (nb : nat) (b : list A) : Queue A :=
  if nf <? nb then MkQueue (nf + nb) (append f (rev b)) 0 []
  else MkQueue nf f nb b.

Definition push {A : Type} (q : Queue A) (x : A) : Queue A :=
  mkQueue (nfront q) (front q) (nback q + 1) (x :: back q).

Definition pop {A : Type} (q : Queue A) : option (A * Queue A) :=
  match front q with
  | x :: f => Some (x, mkQueue (pred (nfront q)) f (nback q) (back q))
  | [] => None
  end.
\end{lstlisting}
\caption{The banker's queue implemented in Gallina.}\label{fig:banker-queue}
\end{figure}

We manually derive all the demand functions of the banker's queue, shown in
\cref{fig:pushD}. The translation is mostly based on our demand semantics, but
we made certain simplifications---we discuss how we validated our manual
translation at the end of this section.

\begin{figure}[t]
\begin{lstlisting}[language=Coq,numbers=left,xleftmargin=4.0ex]
Record QueueA (A : Type) : Type := MkQueueA
  { nfrontA : nat
  ; frontA : T (listA A)
  ; nbackA : nat
  ; backA : T (listA A)
  }.

Definition pushD {A} (q : Queue a) (x : A)
                     (outD : QueueA A) : Tick (T (QueueA A) * T A) :=
  tick >> let+ (frontD, backD) := mkQueueD (nfront q) (front q)
                                   (S (nback q)) (x :: back q) outD in
  ret (Thunk (MkQueueA (nfront q) frontD (nback q) (tailX backD)), Thunk x).

Definition popD {A} (q : Queue A)
                    (outD : option (T A * T (QueueA A))) : Tick (T (QueueA A)) :=
  tick >>
  match front q, outD with
  | [], _ => Tick.ret (exact q) 
  | x :: f, Some (xA, pop_qA) =>
    let+ (fD, bD) := thunkD (mkQueueD (pred (nfront q)) f
                                      (nback q) (back q)) pop_qA in
    ret (Thunk (MkQueueA (nfront q) (Thunk (ConsA xA fD)) (nback q) bD))
  | _, _ => bottom
  end.
\end{lstlisting}
\caption{Demand functions of the banker's
  queue.}\label{fig:pushD}\label{fig:popD}
\end{figure}

We show the cost theorems we have proven for the banker's queue in
\cref{fig:costspec}. We define the \potential{}~(lines 1--2) of the queue to be
twice the length of the \emph{demand} of the queue's front list's minus the
length of its back list, as described in Equality~(\ref{eqn:potential}). The
function \cdc{sizeX} measures the ``length'' of a demand list. It is
parameterized by a natural number which represents how long we consider the
length of \cdc{NilA}, \ie~\cdc{sizeX 0} means that a demand list of \cdc{NilA}
has a length of \cdc{0}. We do need a max function to make sure that the
potential is at least 0, because the type of the \cdc{potential} function
specifies that the potential must be a natural number.

\begin{figure}[t]
\begin{lstlisting}[language=Coq,numbers=left,xleftmargin=4.0ex]
Definition potential (q: QueueA A) : nat :=
  2 * (sizeX 0 (frontD q) - nbackD q).

Definition const : nat := 7.

Theorem pushD_cost {A} (q : Queue A) (x : A) (OutD : QueueA A) :
  well_formed q ->
  qOutD `is_approx` push q x ->
  let (cost, qInD) := pushD q x qOutD in
  potential qInD + cost <= potential qOutD + const.

Lemma popD_cost {A} (q : Queue A) (outD : option (T A * T (QueueA A))) :
  well_formed q ->
  outD `is_approx` pop q ->
  let qA := Tick.val (popD q outD) in
  let cost := Tick.cost (popD q outD) in
  potential qA + cost <= const + potential outD.

Lemma pushD_spec {A} (q : Queue A) (x : A) (outD : QueueA A) :
  outD `is_approx` push q x ->
  forall qD xD, (qD, xD) = Tick.val (pushD q x outD) ->
  let dcost := Tick.cost (pushD q x outD) in
  pushA qD xD [[ fun out cost => outD `less_defined` out /\ cost <= dcost ]].

Lemma popD_spec {A} (q : Queue A) (outD : option (T A * T (QueueA A))) :
  outD `is_approx` pop q ->
  forall qD, qD = Tick.val (popD q outD) ->
  let dcost := Tick.cost (popD q outD) in
  popA qD [[ fun out cost => outD `less_defined` out /\ cost <= dcost ]].
\end{lstlisting}
\caption{Cost specification for the banker's queue.}\label{fig:costspec}
\end{figure}

We then define the cost specification of \cdc{pushD} as an inequality between
the \potential{} of its input demand~(\cdc{qInD}) plus the cost~(\cdc{cost}) and
the \potential{} of its output demand~(\cdc{qOutD}) plus a
constant~(\cdc{const})~(lines 6--10). This is the same as
Inequality~(\ref{eqn:costspec}) but we change the inequality to only include the
\cdc{+} operation so that we only need to use natural numbers in the
specification. The theorem additionally requires the queue \cdc{q} to be well
formed~(\cdc{well_formed}, line 7), which is the invariant that all the banker's
queue's operations maintain: (1)~the \cdc{back} list is not longer than the
\cdc{front} list, and (2)~the \cdc{nfront} and \cdc{nback} fields correctly
represent the lengths of the \cdc{front} list and the \cdc{back} list,
respectively. Similarly, we prove a cost theorem for \cdc{popD}~(lines~12--17).

However, we do not wish to trust our demand semantics, as functions in the
banker's queue are more complicated than insertion sort or selection sort. To
show that our mechanized cost analysis is correct, we additionally translate the
banker's queue using another model of laziness, namely the clairvoyant
semantics~\citep{clairvoyant, forking-paths}. The advantage of a clairvoyant
semantics is that its translation from pure functions is more mechanical: it can
be done by adding the right combinators to the pure function.

We show that the demand functions of the banker's queue agree with the
clairvoyant translation on computation cost in lines 19--29 in
\cref{fig:costspec}. Theorems \cdc{pushD_spec} and \cdc{popD_spec} state the
equivalence relation between the demand functions \cdc{pushD} and \cdc{popD}
with the clairvoyant functions \cdc{pushA} and \cdc{popA}. The \cdc{[[...]]}
notation is called an \emph{optimistic specification}~\cite{forking-paths},
which is a form of incorrectness logic~\citep{incorrectness} that shows the
\emph{existence} of an output approximation and computation cost that satisfies
the property specified in \cdc{[[...]]}. \ifextended We use the optimistic
specification because the clairvoyant semantics is nondeterministic. An
over-approximation will specify all its nondeterministic branches, which
includes branches with unnecessary computations. The optimistic specification,
which is based on under-approximations, specifies the existence of certain costs
on some branches. The optimistic specification helps us specify a more precise
upper bound. \fi

\subsection{Implicit Queue}\label{subsec:implicit}

The implicit queue is another persistent data structure that exhibits amortized
constant computation cost shown by \citeauthor{purely-functional}. We show key
Gallina definitions of the implicit queue in \cref{fig:implicit-queue}.

\begin{figure}
\begin{lstlisting}[language=Coq,numbers=left,xleftmargin=4.0ex]
Inductive Queue (A : Type) : Type :=
| Nil : Queue A
| Deep : Front A -> Queue (A * A) -> Rear A -> Queue A.
    
Fixpoint push (A : Type) (q : Queue A) (x : A) : Queue A :=
  let '(f, m, r) :=
    match q with
    | Nil => (FOne x, Nil, RZero)
    | Deep f m r =>
        let (m, r) :=
          match r with
          | RZero => (m, ROne x)
          | ROne y => (push m (y, x), RZero)
          end in
        (f, m, r)
    end in
  Deep f m r.

Fixpoint pop (A : Type) (q : Queue A) : option (A * Queue A) :=
  match q with
  | Nil => None
  | Deep f m r =>
      let (x, q) :=
        match f with
        | FOne x =>
            let q :=
              match (pop m) with
              | Some yzm' =>
                  let '((y, z), m') := yzm' in
                  Deep (FTwo y z) m' r
              | None =>
                  match r with
                  | ROne y => Deep (FOne y) Nil RZero
                  | RZero => Nil
                  end
              end in (x, q)
        | FTwo x y => (x, Deep (FOne y) m r)
        end in
      Some (x, q)
  end.
\end{lstlisting}
\caption{The Gallina implementation of the implicit queue. Compared to the
  implementation presented by \citet{purely-functional}, we slightly simplify
  the base case to make it only represent an empty queue. This does not affect
  the computation cost of the implicit queue. We also simplify the code for
  readability. The actual code in our artifact is written in ANF for an easier
  translation.}\label{fig:implicit-queue}
\end{figure}

An implicit queue of type \cdc{A} is an inductive data type that is either an
empty queue \cdc{Nil}~(line~2) or a ``deep'' structure~(lines~3) that contains:
(1)~a \cdc{Front}, which contains one or two \cdc{A}s, (2)~a \cdc{Rear}, which
contains zero or one \cdc{A}, and (3)~an inner queue of a product \cdc{A * A}.

When \cdc{push}ing a new element to the queue, we first perform a pattern match
on the input queue \cdc{q}~(line~7). If \cdc{q} is empty, we add the new element
to \cdc{Front} directly~(line~8). If \cdc{q} is a \cdc{Deep} structure~(line~9),
we pattern match on its \cdc{Rear}~(line~11). If its \cdc{Rear} contains zero
elements, we put the new element in its \cdc{Rear}~(line~12). If \cdc{q} is a
\cdc{Deep} structure that already has an element \cdc{y} in the \cdc{Rear}~(the
maximal number of element in \cdc{Rear}), we make a \emph{polymorphic recursive
  call} to \cdc{push} the product of \cdc{y} and the new element \cdc{x} to the
inner queue~(line~13). The use of polymorphic recursion is important for the
efficiency of the implicit queue---the technique is known as \emph{recursive
  slowdown}.

When \cdc{pop}ing an element from the queue, we again first perform a pattern
matching on \cdc{q}~(line~20). When \cdc{q} is \cdc{Nil}, there is simply
nothing to \cdc{pop} so we return \cdc{None}~(line~21). When \cdc{q} is a
\cdc{Deep} structure, we first check its \cdc{Front}. If the \cdc{Front}
contains only one element~(the minimal number of elements in \cdc{Front}), we
\cdc{pop} two elements from the inner queue to make the new \cdc{Front} if the
inner queue still contains some elements~(lines~27--30). If there is no more
element from the inner queue~(line~31), we move the \cdc{Rear} element to
\cdc{Front} if there is one~(line~33) or \cdc{Nil} otherwise (line~34). However,
if the \cdc{Front} contains two elements, we just need to retrieve the first
one~(line~37).

We carefully defined the \cdc{push} and \cdc{pop} functions for the implicit
queue in \cref{fig:implicit-queue} so that they can take advantage of lazy
\cdc{let}-bindings in lazy languages. For example, the \cdc{push} function can
return a \cdc{Deep} structure without evaluating the input queue. Similarly, the
\cdc{push} function can evaluate the \cdc{Front} elements in a \cdc{Deep}
structure without evaluating the \cdc{Rear} of the input queue.

We define demand functions \cdc{pushD} and \cdc{popD} of the implicit queue. One
challenge with these demand functions is polymorphic recursion. For example, the
\cdc{push} function over a polymorphic type \cdc{A} runs a recursive call on the
type \cdc{A * A}~(line~13). In a lazy setting, the type \cdc{A * A} can be half
evaluated---it might have an evaluated value as its first element while its
second element is unevaluated. This requires the demand functions to be even
``more polymorphic'' because it needs to keep track of both pure values and
demands. In practice, we define two extra demand functions \cdc{pushD'} and
\cdc{popD'} that are the result of demand translation of \cdc{push} and
\cdc{pop} but with one more polymorphic type as their parameters. We then define
\cdc{pushD} and \cdc{popD} in terms of \cdc{pushD'} and \cdc{popD'}. We show the
approximation of the implicit queue's \cdc{Queue} datatype, the type signatures
of \cdc{pushD'} and \cdc{popD'}, and the definitions of \cdc{pushD} and
\cdc{popD} in \cref{fig:implicitD}. Interested readers can find detailed
definitions of these functions, which are too long to be shown in this paper, in
\cref{apx:demand-implicit} and in our artifact.

\begin{figure}[t]
\begin{lstlisting}[language=Coq]
Inductive QueueA (A : Type) : Type :=
| NilA : QueueA A
| DeepA : T (FrontA A) -> T (QueueA (prodA A A)) -> T (RearA A) -> QueueA A.
    
Definition pushD' {A B : Type} `{Exact A B}
  (q : Queue A) (x : A) (outD : QueueA B) : Tick (prodA (QueueA B) B).
    
Definition pushD {A : Type} :
  Queue A -> A -> QueueA A -> Tick (prodA (QueueA A) A) :=
  pushD'.

Definition popD' {A B : Type}
  (q : Queue A) (outD : option (T (prodA B (QueueA B)))) : Tick (T (QueueA B)).

Definition popD {A : Type} :
  Queue A -> option (T (prodA A (QueueA A))) -> Tick (T (QueueA A)) :=
  popD'.
\end{lstlisting}
\caption{The approximation of the implicit queue's \cdc{Queue} datatype, the
  type signatures of \cdc{pushD'} and \cdc{popD'}, and the definitions of
  \cdc{pushD} and \cdc{popD} of the implicit queue.}\label{fig:implicitD}
\end{figure}

We also proved all the cost theorems for these demand functions similar to the
banker's queue, which include that the demand functions agree with a clairvoyant
translation of \cdc{push} and \cdc{pop} on computation cost. The cost theorems
that we have proven can be found in \cref{apx:cost-theorems} and in our
artifact. At last, we proved that the implicit queue satisfy
\cdc{Physicist'sArgumentD} and
\cdc{AmortizedCostSpec}~(\cref{fig:coq-physicist}), which means that the
implicit queue is both amortized and persistent.

\section{Related Work}\label{sec:related-work}

\paragraph{Verification of lazy functional programs}
Analyzing computation cost of lazy languages usually requires modeling and
reasoning about mutable heaps~\citep{launchbury1993}, which is challenging.
Prior works on formalized cost analysis of lazy programs can be divided into
three categories. The first approach uses a tick monad~\citep{danielsson-08} to
model computation cost in a functional way. To model sharing, a \texttt{pay}
combinator needs to be manually inserted at proper places to preemptively force
the computation of the shared part of a data structure.
\LiquidHaskell~\citep{liquidate, liquidhaskell-dissertation} uses a similar
approach.

Alternatively, one can reason about mutable heaps using separation logics. This
approach is taken by \citet{iris-thunk} to verify an OCaml implementation of the
lazy banker's queue, the physicist’s queue, and implicit queues in Coq based on
the \irisc framework~\citep{iris-credit}. \Citeauthor{iris-thunk}'s method is
based on an imperative style, as their code directly encodes thunks in OCaml and
their reasoning is based on the Iris separation logic~\citep{iris, iris4}.
However, what they expose to the users are abstract and in the style of
\citeauthor{purely-functional}'s debit-based reasoning with no mutation or
reasoning about ownership involved.

Instead of dealing with the complexity caused by mutable heaps directly, the
third approach uses alternative semantics models instead of natural
semantics~\citep{launchbury1993}. Prior works on \emph{clairvoyance semantics}
fall in this category~\citep{clairvoyant, forking-paths}. The key idea of the
clairvoyance semantics is simulating laziness using a nondeterministic
\emph{call-by-value} model. Our work is based on \citet{bjerner1989}'s demand
semantics, which is untyped and relies on partial functions, whereas our version
is typed, allowing the semantics to be defined in terms of total functions,
making the semantics simpler to formalize and use in a proof assistant based on
type theory such as Coq. We have also formally proved a correspondence between
the demand semantics and a preexisting model of laziness, namely the clairvoyant
semantics, which prior work~\citep{clairvoyant,forking-paths} has connected to
the natural semantics of~\citet{launchbury1993}. Compared with the clairvoyant
semantics, demand semantics allows us to sidestep the need for both
nondeterminism and the need for incorrectness logic~\citep{forking-paths}, by
including output demands as parts of input for the demand functions.

One drawback of our method is that, due to changing ``when'' a computation
happens in our model, our method cannot be used to analyze real-time computation
cost, nor can it be used to analyze other types of resource that are not
monotone, \eg~memory usage. In comparison, \LiquidHaskell can be used on
measuring the usage of ``any kind of resource whose usage is
additive''~\citep{liquidate}. \irisc has been used on verifying the implicit
queue that has real-time constant computation cost.

On the testing side, \citet{foner2018} introduce a low level Haskell testing
library \texttt{StrictCheck} using a similar idea of demand-driven analysis.
They focus on testing for strictness bugs in lazy programs, utilizing persistent
and non-persistent queues introduced by \citeauthor{purely-functional} as
examples to verify their implementation, much as we have.

\paragraph{Demand-driven program analysis and symbolic execution}
Demand-driven analysis is also a useful technique that has been used in
data-flow analysis, control-flow analysis, and symbolic
execution~\citep{horwitz95, demand-adpative, ddpa, ddse, germane19}. The key
idea is by starting from the target to be analyzed, unneeded analysis/evaluation
can be avoided. Our demand semantics is based on similar idea, as it avoids
nondeterminism present in clairvoyant semantics by having output demand as parts
of demand function's input. However, our demand semantics focuses on reasoning
about computation cost for lazy functional programs and lazy, amortized, and
persistent data structures. \Citet{ddpa, ddse, germane19} also support program
analysis and symbolic execution for \emph{higher-order} functions, which our
demand semantics does not currently support.

\paragraph{Demand analysis in compilers}
Compilers like the Glasgow Haskell Compiler~(GHC) employs demand analysis to
perform compiler optimizations~\citep{demand-haskell, cardinality-analysis}.
However, demand analyses in compilers typically focus on finding one-shot
lambdas, single-entry thunks, \etc\ to apply optimizations such as
deforestation. Our demand semantics focuses on compositional mechanized
reasoning for proving properties about computation cost.

\paragraph{Cost analysis or amortized analysis for non-lazy semantics}
Cost analysis or amortized cost analysis in the context of strict semantics has
been the subject of much research as well~\citep{static-cost-danner,
  coq-internal-running-times, cutler20, raml, lambda-amor}. \Citet{cutler20}
provide a formalized framework for reasoning about amortized analysis in an
imperative setting. \Citet{lambda-amor} introduce a type system which can embed
call-by-value and call-by-name evaluation as well as accounting for cost savings
via amortization. \Citet{raml} improve this type of analysis by introducing
arbitrary multivariate polynomial functions to their cost representations,
including comparisons of theoretical bounds to real-world examples. All of these
models allow for space-usage analysis, but none of them support reasoning about
laziness.

Calf is a cost-aware logical framework for reasoning about resource usage in
full-spectrum dependently-typed functional programs~\citep{calf}. The language
also supports effects via the call-by-push-value evaluation~\citep{cbpv,
  cbpv-dt}. The framework has been utilized to reason about amortized cost via
\emph{coinduction}~\citep{calf-coind}.

\paragraph{Compiler optimization for lazy functional languages} Compilers for
lazy functional languages employ techniques such as strictness analysis to avoid
creating unnecessary thunks~\citep{strictness-analysis,
  proj-strictness-analysis}. \Citet{optimistic-eval} experimented with a more
aggressive optimization called ``optimistic evaluation'' that speculatively
evaluates thunks but aborts if the compiler decides that it's a bad choice.
Their results show significant improvement over GHC, but it was ultimately not
incorporated in GHC due to its
complexity.\footnote{\url{https://mail.haskell.org/pipermail/haskell/2006-August/018424.html}}

\paragraph{Improvement Theory}
Improvement theory studies if one program \emph{improves} another program in
terms of reduction steps under all program contexts~\citep{improve-cbn,
  improve-cbv}. \Citet{lazy-improve} showed that the improvement theory can be
applied in a lazy setting as well, based on prior work on the call-by-need
$\lambda$-calculus~\citep{lazy-calculus}. \Citet{call-by-need-space-improve}
also studied the space complexity under lazy evaluation using the improvement
theory. Our work focuses on computation cost and amortized cost analysis based
on proof assistants, rather than improvement relations between programs.

\paragraph{Machine-checked complexity theory}
\Citet{forster2017} formalize a weak call-by-value $\lambda$-calculus in Coq
called \emph{L}. They have used L as a model of computation to formally prove
the Cook-Levin Theorem~\citep{gaher_l, cook-levin}.

\section{Conclusion and Future Work}\label{sec:conclusion}

In this paper, we present a demand semantics for lazy functional programs that,
given any valid output demand, returns the minimal input demand required. We
base our demand semantics on \citet{bjerner1989}, but we expand it to support
higher-order functions such as $\mathrm{foldr}$. In addition, we formally prove
that the demand semantics is equivalent to the natural semantics of laziness, by
showing that it is equivalent to the clairvoyant semantics.

We demonstrate the effectiveness of our approach by applying our method to
formally prove that \citeauthor{purely-functional}'s banker's queue is amortized
and persistent using the Coq theorem prover. In the process, we propose a novel
\method{} that allows reasoning about amortization and persistence based on the
demand semantics in a modular way.

In future work, we would like to apply this approach to larger lazy functional
data structures such as finger trees~\citep{finger-trees, finger-trees-pearl}.
We would also like to develop a tool that automatically generates a function's
demand function based on the demand semantics. By integrating this translator
with tools like \hstocoq, we can apply our method to more functional programs
and data structures in mainstream languages such as Haskell.

\section*{Acknowledgment}

We thank all the ICFP 2024 reviewers and PLDI 2024 reviewers whose suggestions
helped significantly improve this paper. We thank Joseph W.\ Cutler and Cassia
Torczon for their involvement during the early stage of this project. Katie
Casamento, James Hook, Mark P.\ Jones, Allison Naaktgeboren, Andrew Tolmach, and
Grant VanDomelen have provided valuable suggestions during the authors'
discussions with them. We also appreciate the feedback from participants at UCSC
LSD Seminar and PNW PLSE Workshop 2024. This work was partially supported by the
\grantsponsor{}{National Science Foundation}{} under Grant
Nos.~\grantnum{}{CCF-2006535} and ~\grantnum{}{CNS-2244494}.

\appendix

\section{Appendix: Rocq Prover Formalization of the Implicit Queue}

\subsection{Demand Functions}\label{apx:demand-implicit}

\begin{lstlisting}[language=Coq]
Inductive QueueA (A : Type) : Type :=
| NilA : QueueA A
| DeepA : T (FrontA A) -> T (QueueA (prodA A A)) -> T (RearA A) -> QueueA A.
  
Fixpoint pushD' (A B : Type) `{Exact A B} (q : Queue A) (x : A) (outD : QueueA B) :
  Tick (prodA (QueueA B) B) :=
  let+ qD :=
    Tick.tick >>
      match outD with
      | DeepA fD mD rD =>
          match q with
          | Nil => Tick.ret (Thunk NilA)
          | Deep f m r =>
              match r with
              | RZero => Tick.ret (Thunk (DeepA fD mD (Thunk RZeroA)))
              | ROne y =>
                  let+ uD := thunkD (pushD' m (y, x)) mD in
                  let '(pairA mD pD) := uD in
                  let (yD, xD) :=
                    match pD with
                    | Thunk (pairA yD xD) => (yD, xD)
                    | _ => bottom
                    end in
                  Tick.ret (Thunk (DeepA fD mD (Thunk (ROneA yD))))
              end
          end
      | _ => bottom
      end in
    Tick.ret (pairA qD (exact x)).

Definition pushD (A : Type) : Queue A -> A -> QueueA A -> Tick (prodA (QueueA A) A) :=
  pushD'.

Fixpoint popD' (A B : Type) (q : Queue A) (outD : option (T (prodA B (QueueA B)))) :
  Tick (T (QueueA B)) :=
  Tick.tick >>
    match q with
    | Nil => Tick.ret (Thunk NilA)
    | Deep f m r =>
        let+ (fD, mD, rD) :=
          let (xD, qD) :=
            match outD with
            | Some (Thunk (pairA xD qD)) => (xD, qD)
            | _ => bottom
            end in
          match f with
          | FOne x =>
              let p := pop m in
              let (pD, rD) :=
                match p with
                | Some (yz, m') =>
                    match qD with
                    | Thunk (DeepA fD mD' rD) =>
                        let yzD :=
                          Thunk (match fD with
                                 | Thunk (FTwoA yD zD) => pairA yD zD
                                 | _ => bottom
                                 end) in
                        (Thunk (Some (Thunk (pairA yzD mD'))), rD)
                    | _ => bottom
                    end
                | None =>
                    let rD :=
                      match r with
                      | RZero => Thunk RZeroA
                      | ROne y =>
                          let yD :=
                            match qD with
                            | Thunk (DeepA (Thunk (FOneA yD)) _ _) => yD
                            | _ => bottom
                            end in
                          Thunk (ROneA yD)
                      end in
                    (Thunk None, rD)
                end in
              let+ mD := thunkD (popD' m) pD in
              Tick.ret (Thunk (FOneA xD), mD, rD)
          | FTwo x y =>
              let '(yD, mD, rD) :=
                match qD with
                | Thunk (DeepA fD' mD rD) =>
                    let yD :=
                      match fD' with
                      | Thunk (FOneA yD) => yD
                      | _ => bottom
                      end in
                    (yD, mD, rD)
                | _ => bottom
                end in
              Tick.ret (Thunk (FTwoA xD yD), mD, rD)
          end in
        Tick.ret (Thunk (DeepA fD mD rD))
    end.

Definition popD (A : Type) (q : Queue A) (outD : option (T (prodA A (QueueA A)))) :
  Tick (T (QueueA A)) :=
  popD' q outD.
\end{lstlisting}

\subsection{Potential Functions and Cost Theorems}\label{apx:cost-theorems}

\begin{lstlisting}[language=Coq]
Definition size_FrontA (A : Type) (fA : FrontA A) : nat :=
  match fA with
  | FOneA _ => 1
  | FTwoA _ _ => 2
  end.

Definition size_RearA (A : Type) (rA : RearA A) : nat :=
  match rA with
  | RZeroA => 0
  | ROneA _ => 1
  end.

Instance Potential_QueueA : forall (A : Type), Potential (QueueA A) :=
  fix potential_QueueA (A : Type) (qA : QueueA A) :=
    match qA with
    | NilA => 0
    | DeepA fD mD rD =>
        let c := T_rect _ size_FrontA 2 fD - T_rect _ size_RearA 0 rD
        in c + @Potential_T _ (potential_QueueA _) mD
    end.
  
Lemma pushD'_cost : forall (A B : Type) `{LessDefined B, Exact A B}
  (q : Queue A) (x : A) (outD : QueueA B),
    outD `is_approx` push q x ->
    let inM := pushD' q x outD in
    let cost := Tick.cost inM in
    let (qD, _) := Tick.val inM in
    potential qD + cost <= 2 + potential outD.
    
Corollary pushD_cost : forall (A : Type) `{LessDefined A}
  (q : Queue A) (x : A) (outD : QueueA A),
    outD `is_approx` push q x ->
    let inM := pushD q x outD in
    let cost := Tick.cost inM in
    let (qD, _) := Tick.val inM in
    potential qD + cost <= 2 + potential outD.

Lemma pushD'_spec (A B : Type) :
  forall `{LDB : LessDefined B, !Reflexive LDB, Exact A B}
    (q : Queue A) (x : A) (outD : QueueA B),
    outD `is_approx` push q x ->
    forall qD xD, pairA qD xD = Tick.val (pushD' q x outD) ->
             let dcost := Tick.cost (pushD' q x outD) in
             pushA qD xD [[ fun out cost =>
                              outD `less_defined` out /\ cost <= dcost ]].
    
Corollary pushD_spec (A : Type) :
  forall `{LDA : LessDefined A, !Reflexive LDA}
    (q : Queue A) (x : A) (outD : QueueA A),
    outD `is_approx` push q x ->
    forall qD xD, pairA qD xD = Tick.val (pushD' q x outD) ->
             let dcost := Tick.cost (pushD' q x outD) in
             pushA qD xD [[ fun out cost =>
                              outD `less_defined` out /\ cost <= dcost ]].

Lemma popD'_cost : forall (A B : Type)
                     `{LessDefined B, Exact A B}
                     (q : Queue A) (outD : option (T (prodA B (QueueA B)))),
    outD `is_approx` pop q ->
    let d := match outD with
             | Some (Thunk (pairA _ qD)) => potential qD
             | _ => 0
             end in
    let inM := popD' q outD in
    let cost := Tick.cost inM in
    let inD := Tick.val inM in
    potential inD + cost <= 3 + d.
    
Corollary popD_cost :
  forall (A : Type) `{LessDefined A}
         (q : Queue A) (outD : option (T (prodA A (QueueA A)))),
    outD `is_approx` pop q ->
    let d := match outD with
             | Some (Thunk (pairA _ qD)) => potential qD
             | _ => 0
             end in
    let inM := popD' q outD in
    let cost := Tick.cost inM in
    let inD := Tick.val inM in
    potential inD + cost <= 3 + d.

Lemma popD'_spec :
  forall (A B : Type) `{LDB : LessDefined B, !Reflexive LDB, Exact A B}
    (q : Queue A) (outD : option (T (prodA B (QueueA B)))),
    outD `is_approx` pop q ->
    forall qD, qD = Tick.val (popD' q outD) ->
    let dcost := Tick.cost (popD' q outD) in
    popA qD [[ fun out cost => outD `less_defined` out /\ cost <= dcost ]].

Corollary popD_spec :
  forall (A : Type) `{LDA : LessDefined A, !Reflexive LDA}
    (q : Queue A) (outD : option (T (prodA A (QueueA A)))),
    outD `is_approx` pop q ->
    forall qD, qD = Tick.val (popD' q outD) ->
    let dcost := Tick.cost (popD' q outD) in
    popA qD [[ fun out cost => outD `less_defined` out /\ cost <= dcost ]].
\end{lstlisting}

\bibliography{ref}

\end{document}